\documentclass{aa}

\usepackage{graphicx}
\usepackage{amssymb,amsmath}
\usepackage{color}
\usepackage{epstopdf}
\usepackage{natbib}
\usepackage{multirow}
\usepackage[colorlinks=true,     linkcolor=blue, citecolor=blue, filecolor=blue, urlcolor=blue]{hyperref}
\usepackage{placeins}
\usepackage{orcidlink}

\title{The assembly of the most rotationally supported disc galaxies in the TNG100 simulations} 
\author{
  Silvio Rodr\'iguez\inst{1}\thanks{E-mail:silvio.rodriguez@unc.edu.ar}\orcidlink{0000-0001-5329-5242},
  Valeria A. Cristiani\inst{1}\orcidlink{0000-0003-0143-7573},
  Laura V. Sales\inst{2}\orcidlink{0000-0002-3790-720X} \and
  Mario G. Abadi\inst{1, 3}\orcidlink{0000-0003-3055-6678}
  }
\authorrunning{S. Rodr\'iguez et al.}
\institute{
Instituto de Astronom\'ia Te\'orica y Experimental, UNC-CONICET, C\'ordoba, X5000BGR, Argentina\\
\and
Department of Physics and Astronomy, University of California, Riverside, CA, 92521, USA\\
\and
Observatorio Astron\'omico de C\'ordoba, Universidad Nacional de C\'ordoba, X5000BGR, Argentina
}

\date{Accepted --. Received --; in original form --}

\abstract
{Disc dominated galaxies can be difficult to accommodate in a hierarchical formation scenario like $\Lambda$CDM, where mergers are an important growth mechanism. However, observational evidence indicates that these galaxies are common in our universe.}
{We seek to characterise the conditions that lead to the formation of disc dominated galaxies within $\Lambda$CDM.}
{We use dynamical decomposition of the stellar particles in all galaxies with stellar mass $M_*=[10^{10} \rm - 10^{11}]\; \rm M_\odot$ within the cosmological hydrodynamical simulation Illustris TNG100. We select a sample of 43 mostly-disc galaxies having less than $\sim 10\%$ of their mass into a bulge component. For comparison, we also study two additional stellar-mass matched samples: 43 intermediate galaxies having $\sim 30\%$ of their stellar mass in the bulge and 43 with purely spheroidal-like morphology.}
{We find that the selection purely based on stellar dynamics is able to reproduce the expected stellar population trends of different morphological types, with higher star-formation rates and younger stars in disc-dominated galaxies. Halo spin seems to play no role on the morphology of the galaxies, in agreement with previous works. At fixed $M_*$, our mostly-disc and intermediate samples form in dark matter haloes that are $2$-$10$ times less massive than the spheroidal sample, highlighting a higher efficiency in disc galaxies to retain and condensate their baryons. On average, mergers are less prevalent in the build up of discs than in spheroidal galaxies, but there is a large scatter, including the existence of mostly-disc galaxies with $15\%$-$30\%$ of their stars from accreted origin. Discs start forming early on, settling their low vertical velocity dispersion as early as $9$-$10$ Gyr ago, although the dominance of the disc over the spheroid gets established more recently ($3$-$4$ Gyr lookback time). The most rotationally supported discs form in haloes with the lowest virial mass in the sample and best aligned distribution of angular momentum in the gas.}
{}


\begin{document}
\maketitle
\nolinenumbers

\section{Introduction}
\label{sec:intro}

In a hierarchical formation scenario such as the $\Lambda$ cold dark matter ($\Lambda$CDM) cosmological model, where structures form bottom-up and grow mainly via mergers \citep{White1978, Peebles1982, Blumenthal1984,  Davis1985, Kauffmann1993, Cole2008, Fakhouri2008, Neistein2008}, a high prevalence of pure disc galaxies is unexpected. However, in the local volume, a third to a half of the giant galaxies are reported to have no bulges, at least in their classical definition \citep{Kautsch2006, Kormendy2008, Kormendy2010} and seem mostly supported by stars on circular orbits \citep{Peebles2020}. More representative galaxy samples covering larger volumes also show axis ratios consistent with the presence of razor-thin discs in sky-wide surveys like GAMA, SDSS and SAMI \citep{Haslbauer2022}. How galactic discs can retain their orderly rotation amidst the intrinsically chaotic assembly of mass in  $\Lambda$CDM remains an interesting question.

The existence of velocity-dispersion supported systems, such as galaxy bulges, are comparatively much easier to accommodate within the cosmological assembly of galaxies. For example, mergers are identified as one of the main formation and growth mechanisms of bulges \citep{Toomre1977, Walker1996, Hopkins2012, Martig2012}. Additionally, gravitational instabilities \citep{Noguchi1999, Elmegreen2004} or the accretion of gas on misaligned orbits \citep{Scannapieco2009,Sales2012} may also lead to growth of spheroidal components. Although a high gas fraction \citep{Noguchi1999, Elmegreen2004,Hopkins2009} or well defined co-planar orbits may help preserve or even grow the disc in galaxies \citep{Abadi2003,DiCintio2019}, the factors n\"aively expected to pull the orbits of stars away from circular motion seem by far too abundant in $\Lambda$CDM to easily accommodate the large number of disc dominated galaxies observed. In fact, several authors have pointed out that theoretical expectations for the fraction of stellar mass in discs of simulated galaxies is too low compared to the observed galaxy population \citep[e.g., ][]{Stewart2008, Kormendy2010,  Peebles2020,Haslbauer2022}.

The task of comparing observational results to the morphology predicted in galaxy formation models within $\Lambda$CDM should not be taken lightly as we do not yet fully understand the effects of numerical resolution and simplified assumptions in the treatment of baryons on the final structure of simulated galaxies. For example, \citet{BenitezLlambay2018} argues that the disc thickness depends on both, the sound speed and ability to resolve gravity well (among other factors), which might change with resolution, mass of the galaxy and, most importantly, the baryonic treatment chosen for the simulation. Uncertainties in the modelling of stellar feedback have also been shown to directly affect the morphology of stellar and gas discs \citep{Sales2010,Scannapieco2012,Agertz2016}. In addition, modelling of dust is crude to non-existent in current available simulations, but can strongly impact measures of morphology by changing the light distribution in the inner regions.

In view of these challenges, we steer away from the question of whether current simulations form {\it enough} discs --which has been addressed in some of the papers mentioned above-- and instead chose to focus here on the mechanisms that give rise to the formation of the most rotationally supported discs within a large cosmological simulation. We refer to them as our {\it mostly-disc}  galaxies, with less than $\sim 10\%$ of their mass within a bulge-like component. To provide context, we also select other two control samples with the same number and stellar mass distribution than our mostly-discs galaxies: an intermediate morphology sample with a larger contribution of a bulge ($\sim 30\%$) and a spheroidal galaxy sample (no discs). Note that mostly-disc and intermediate samples are both dominated in mass by the disc component, and we might refer to them collectively with ``disc-dominated" when necessary. Our paper is organised as follows. In Sect. \ref{sec:sample} we introduce our simulated sample. In Sec. \ref{sec:assembly} and \ref{sec:mergers} we explore the formation and assembly histories of those galaxies. In Sec.~\ref{sec:discform} we study the formation and time evolution of the discs. We summarise our main findings in Sec.~\ref{sec:concl}.

\section{Galaxy selection}
\label{sec:sample}

We use data from the Illustris TNG100 hydrodynamical simulation \citep{tng1, tng2, tng3, tng4, tng5}. This simulation has a comoving box size of 75000 $\mbox{kpc}\,h^{-1}$, with $1820^3$ dark matter particles. The average mass of the gas particles is 1.4$\times10^6 \mbox{M}_{\sun}$, while the mass of the dark matter particle is 7.5$\times10^6 \mbox{M}_{\sun}$. The simulation uses cosmological parameters consistent with Planck 2015 cosmology \citep{Planck2015}, including $\Omega_m=0.3089$ and $\Omega_{\Lambda}=0.6911$. The baryonic treatment takes into account the main processes at play in the formation of galaxies including a prescription for heating and cooling of the gas, formation of stars and their evolution, metal enrichment and feedback from stars and black holes. 

\begin{figure}
 \begin{center}
  \includegraphics[width=0.45\textwidth]{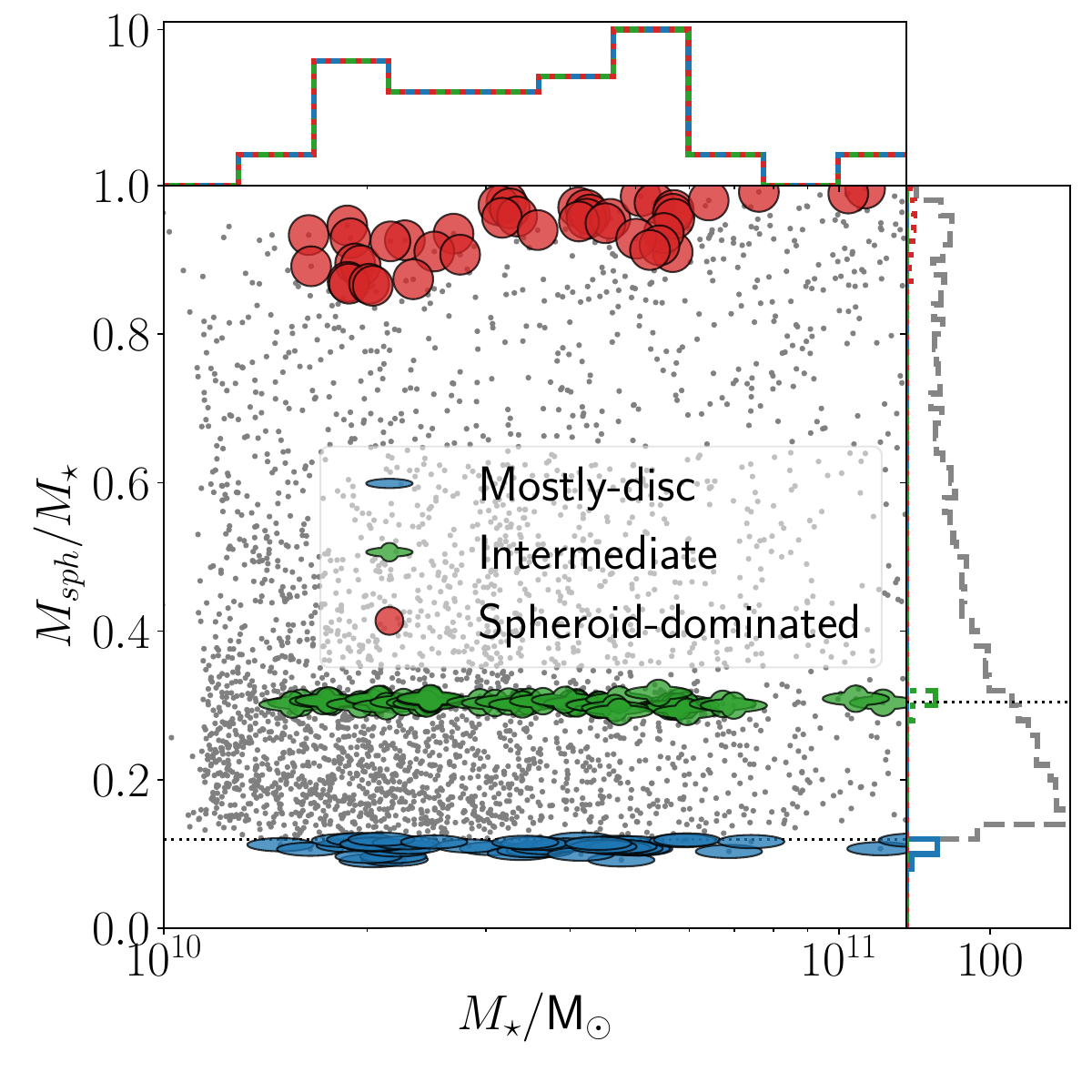}
  \caption{\label{fig_RatioMass} The ratio between the mass of the spheroidal component and the total stellar mass ($M_{sph}/M_{\star}$) as a function of the stellar mass ($M_{\star}$). Gray dots show all the central galaxies with masses over $10^{10} \mbox{M}_{\sun}$ in TNG100. The blue ellipses highlight our population of mostly-disc galaxies, selected to have a negligible spheroidal component, the green ellipses with a circle in the middle indicates our intermediate sample with $M_{sph}/M_{\star}$ value near the median of all the galaxies, and the red circles indicate the spheroid-dominated galaxies selected to have a negligible disc component at a given $M_\star$. The dotted black line indicates the cut used to select the disc dominated sample. The distribution of $M_{\star}$ and $M_{sph}/M_{\star}$ for the selected samples are also shown at the top and left of the panel, respectively (blue solid lines for the mostly-disc dominated sample, green dot-dashed lines for the intermediate sample and dotted red lines for the spheroid-dominated sample), also the distribution of $M_{sph}/M_{\star}$ for all the central galaxies is shown alongside the selected samples with a grey dashed line. The black dotted line in the $M_{sph}/M_{\star}$ distribution indicates the median of $M_{sph}/M_{\star}$ for all the galaxies.
  }
 \end{center}
\end{figure}

From this simulation, we select central galaxies (no satellites) identified by {\sc SUBFIND} \citep{Springel2001, Dolag2009} with stellar masses $M_\star \geq 10^{10} \mbox{M}_{\sun}$. We measure stellar and gas mass ($M_{\rm gas}$) within 3 times the stellar half-mass radius (calculated by {\sc SUBFIND}) at every snapshot. All our galaxies are resolved with $7,000+$ stellar particles, allowing for a robust quantification of their structure. Virial quantities such as virial mass, $M_{200}$, and virial radius, $r_{200}$, are defined at the radius that encloses $200$ times the critical density of the universe. 

\begin{figure*}
 \begin{center}
  \includegraphics[width=0.19\textwidth]{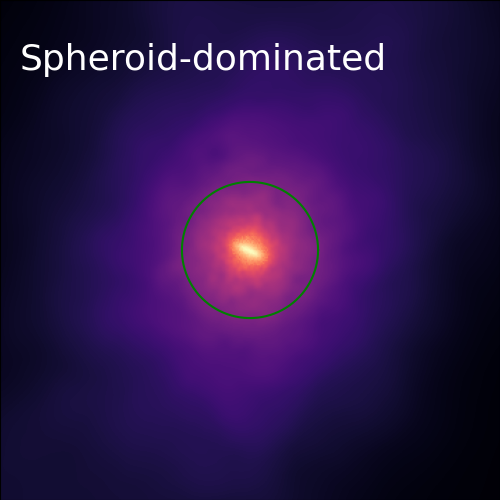}
  \includegraphics[width=0.19\textwidth]{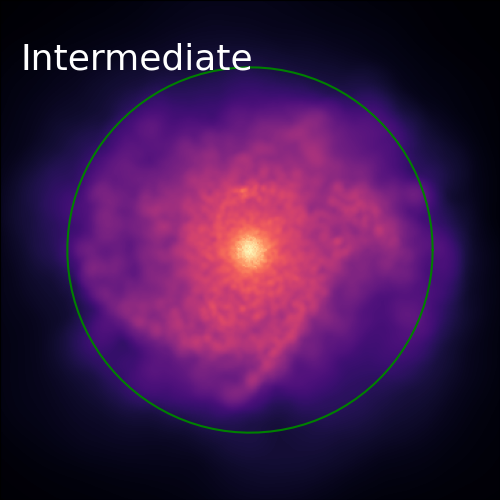}
  \includegraphics[width=0.19\textwidth]{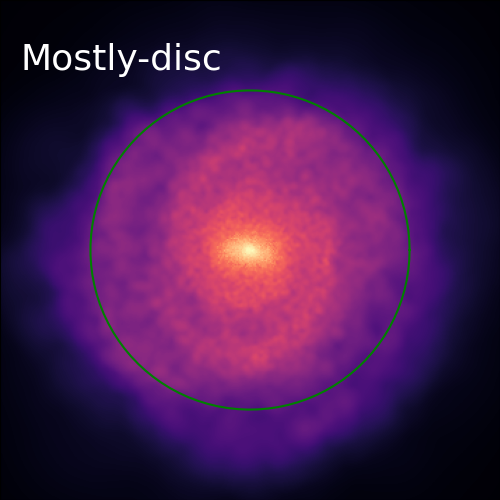}
  \includegraphics[width=0.19\textwidth]{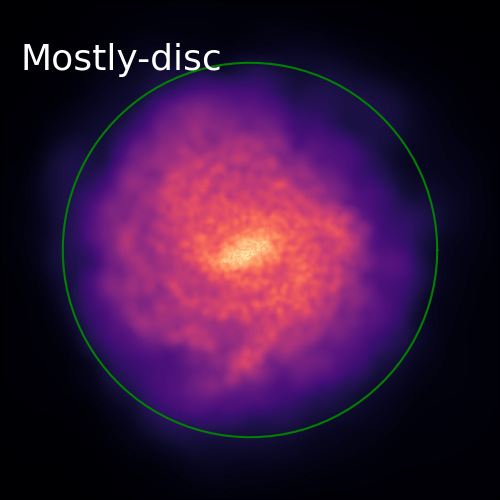}
  \includegraphics[width=0.19\textwidth]{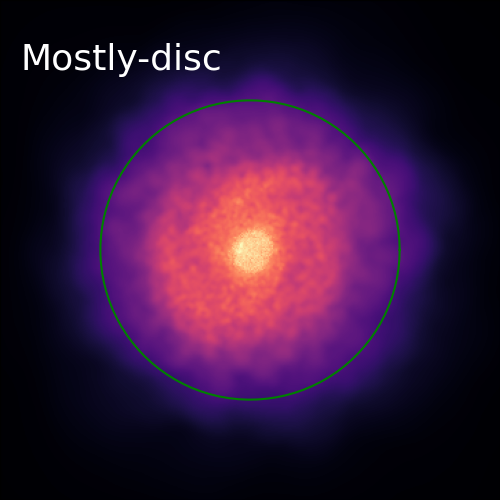}

  \includegraphics[width=0.19\textwidth]{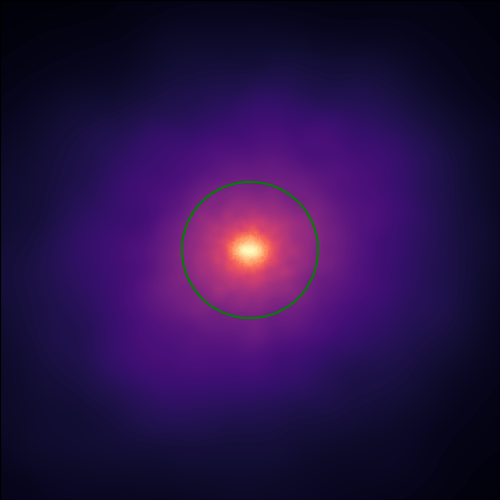}
  \includegraphics[width=0.19\textwidth]{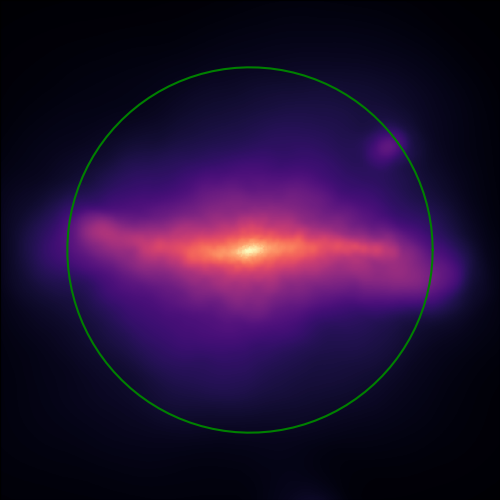}
  \includegraphics[width=0.19\textwidth]{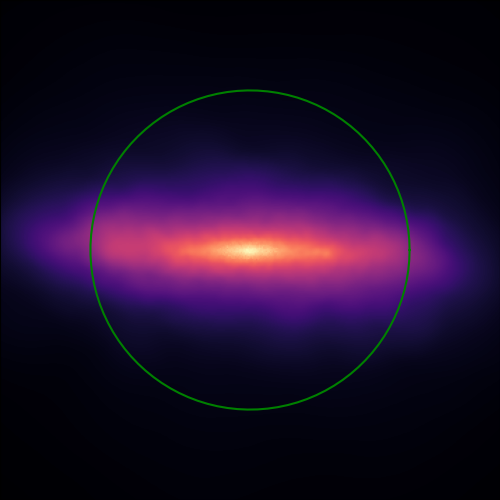}
  \includegraphics[width=0.19\textwidth]{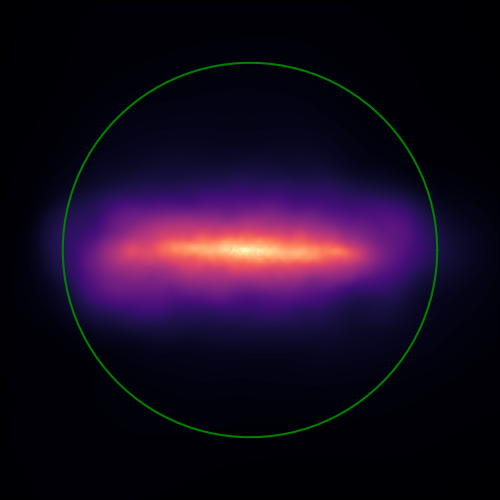}
  \includegraphics[width=0.19\textwidth]{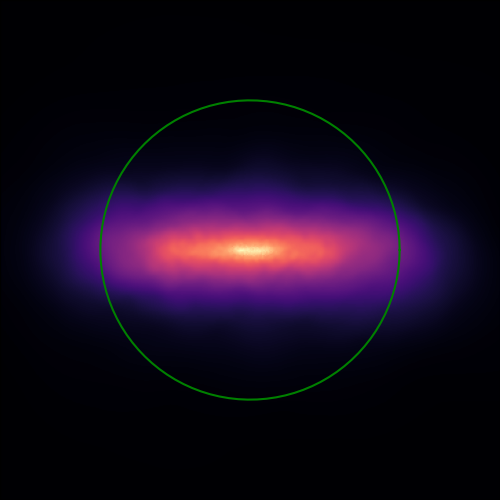}

  \includegraphics[width=0.19\textwidth]{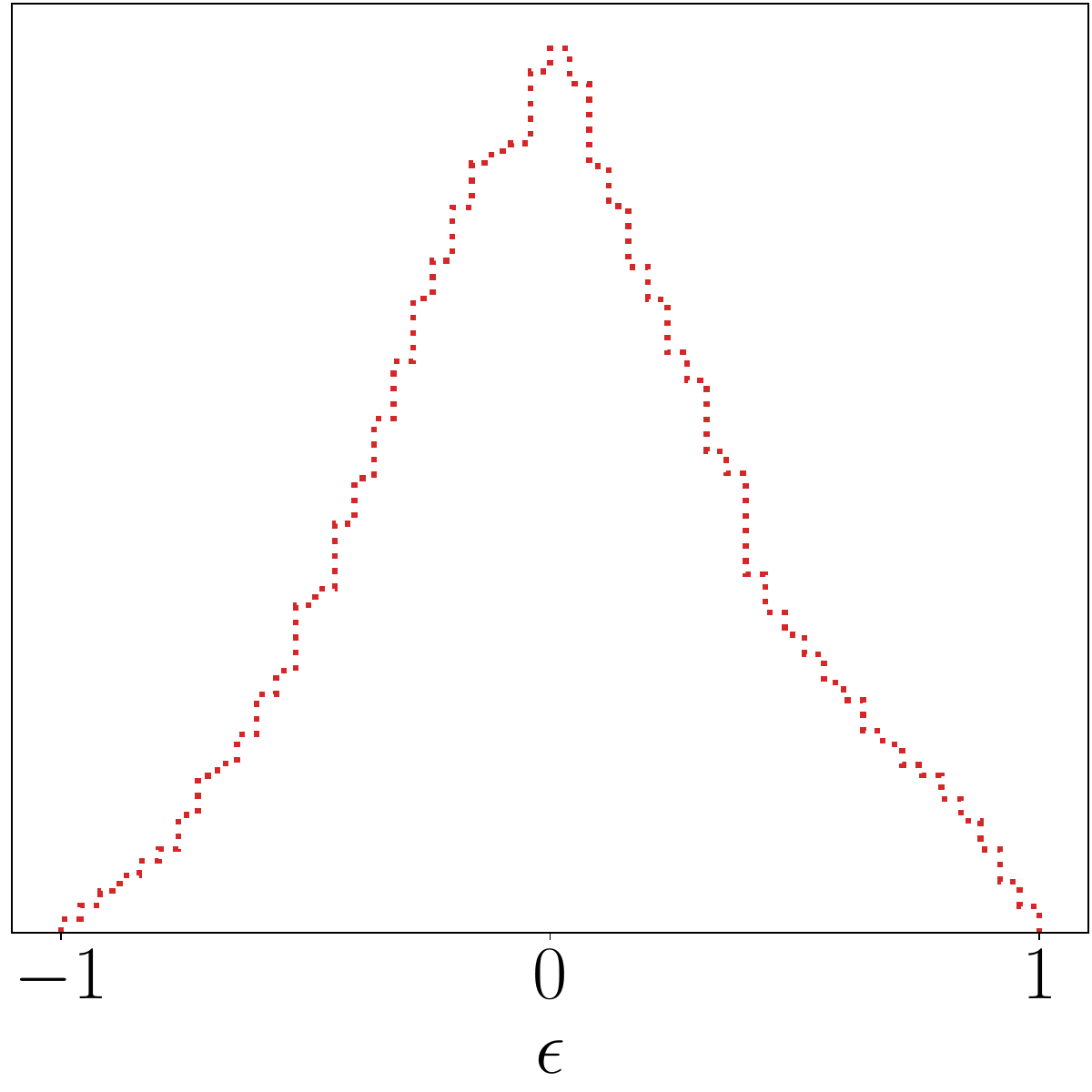}
  \includegraphics[width=0.19\textwidth]{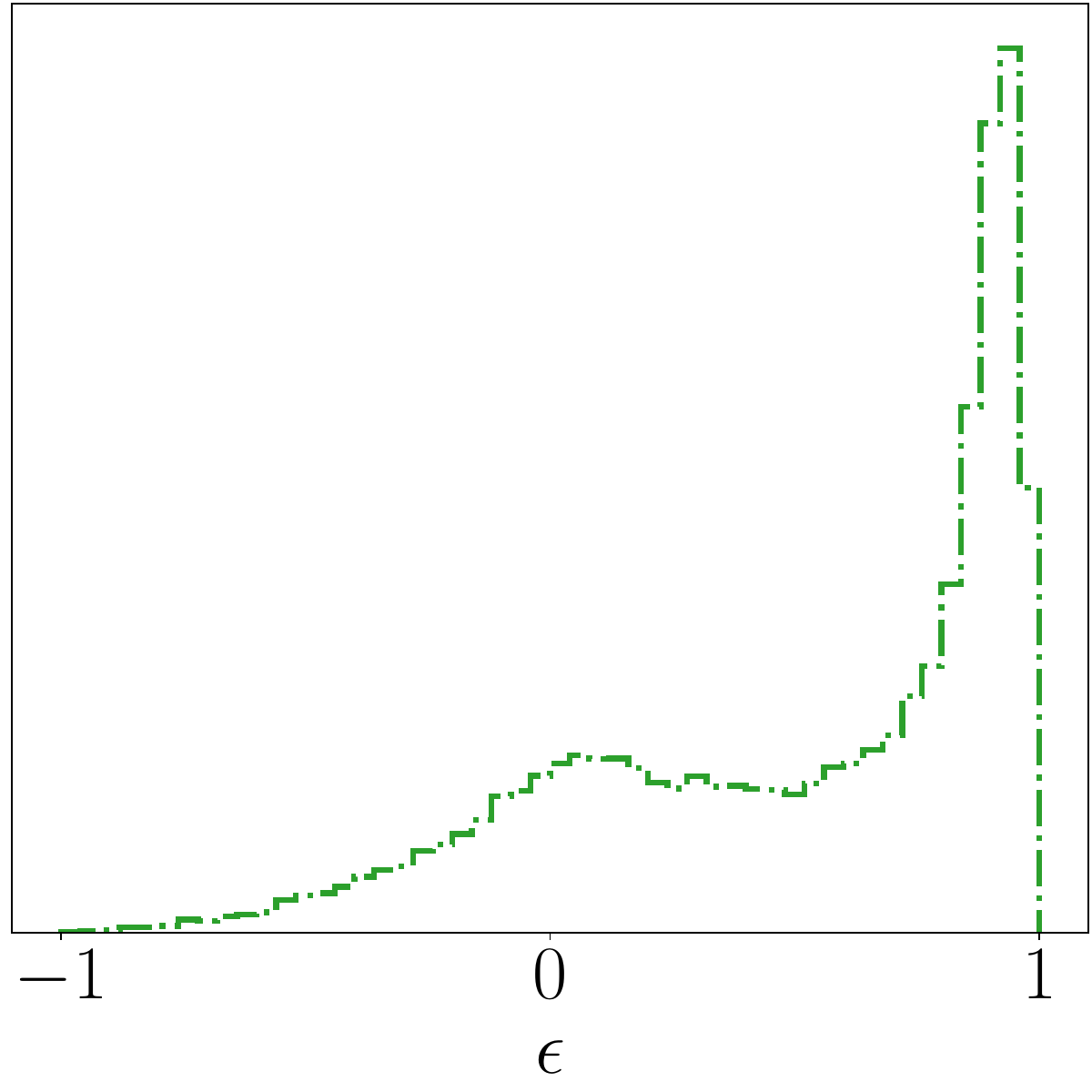}
  \includegraphics[width=0.19\textwidth]{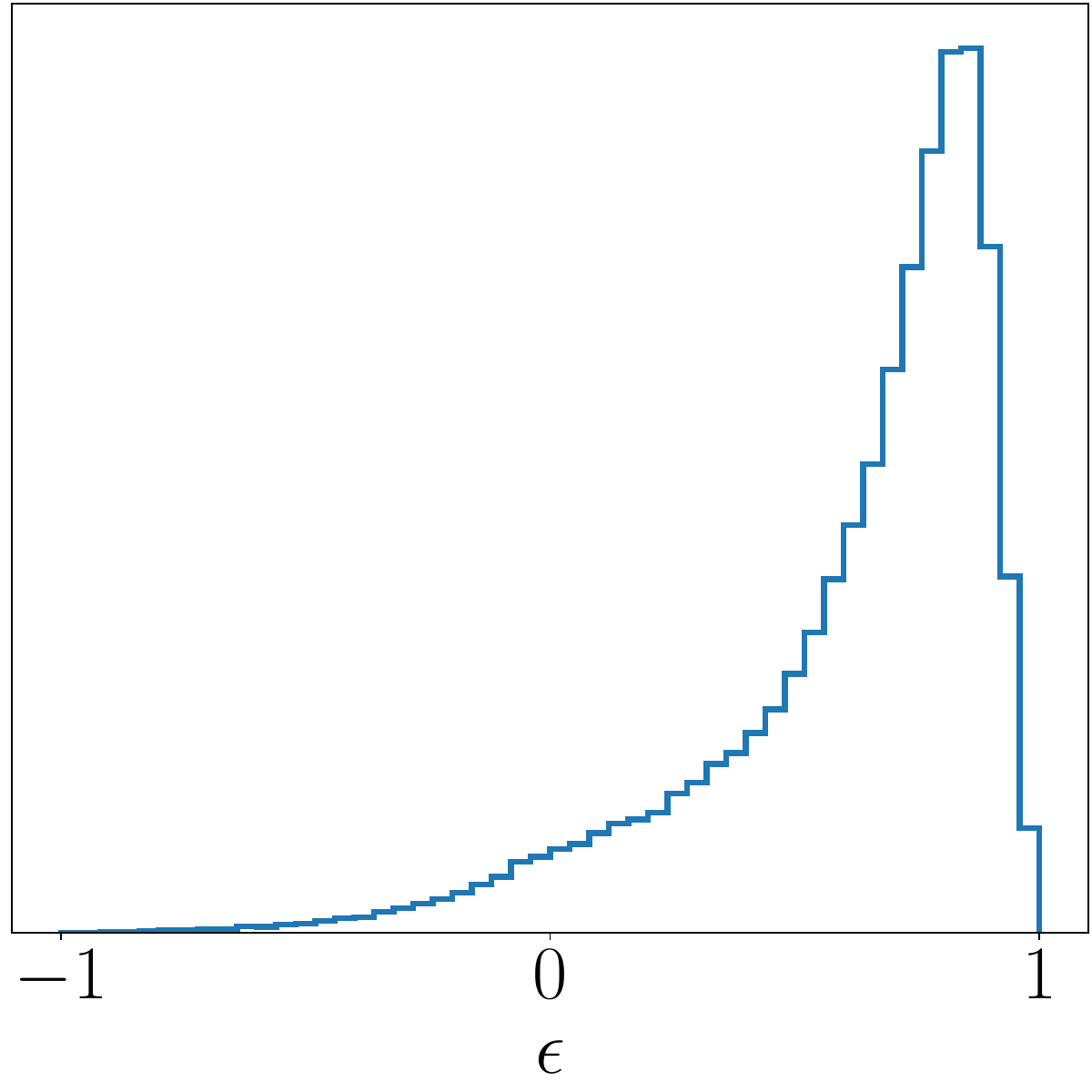}
  \includegraphics[width=0.19\textwidth]{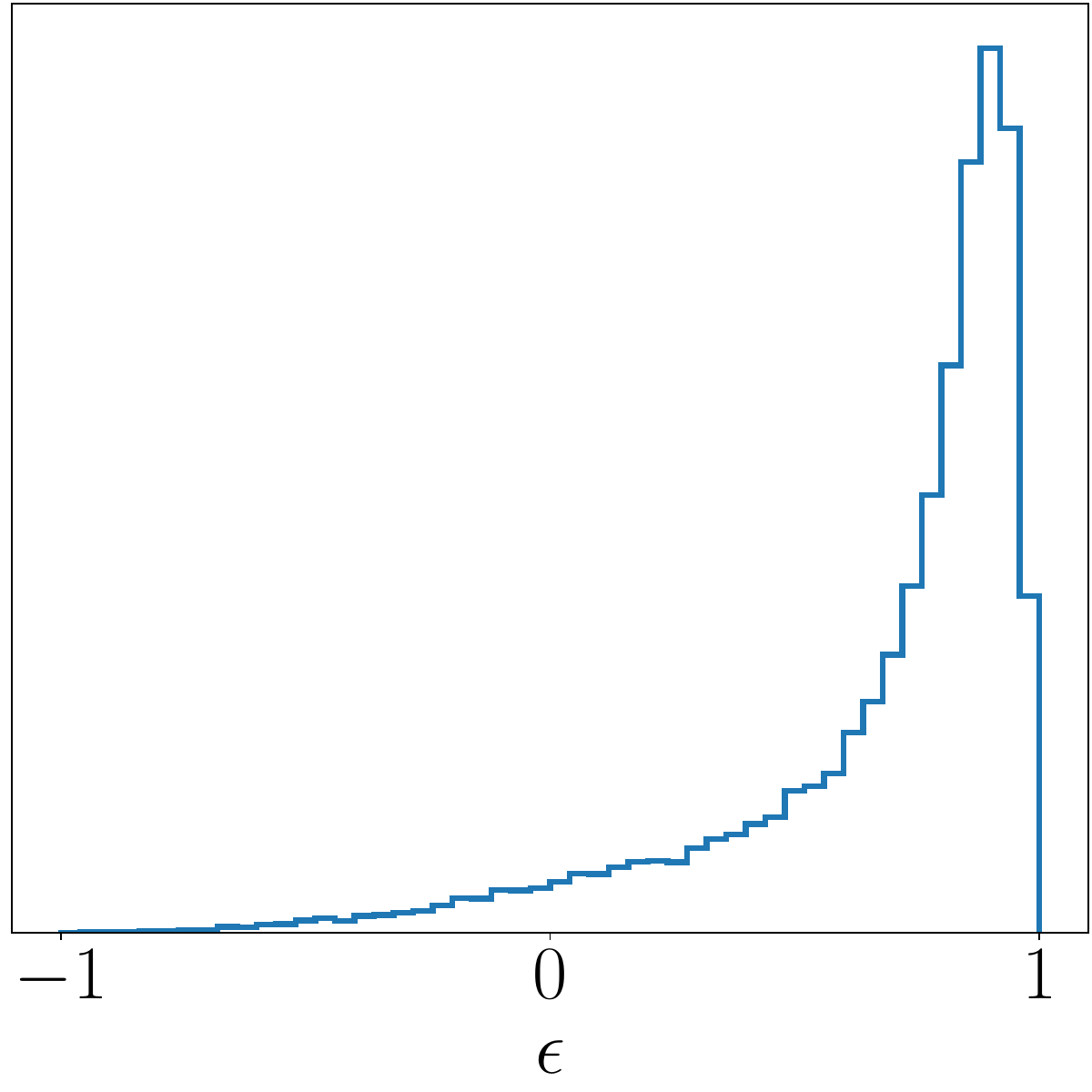}
  \includegraphics[width=0.19\textwidth]{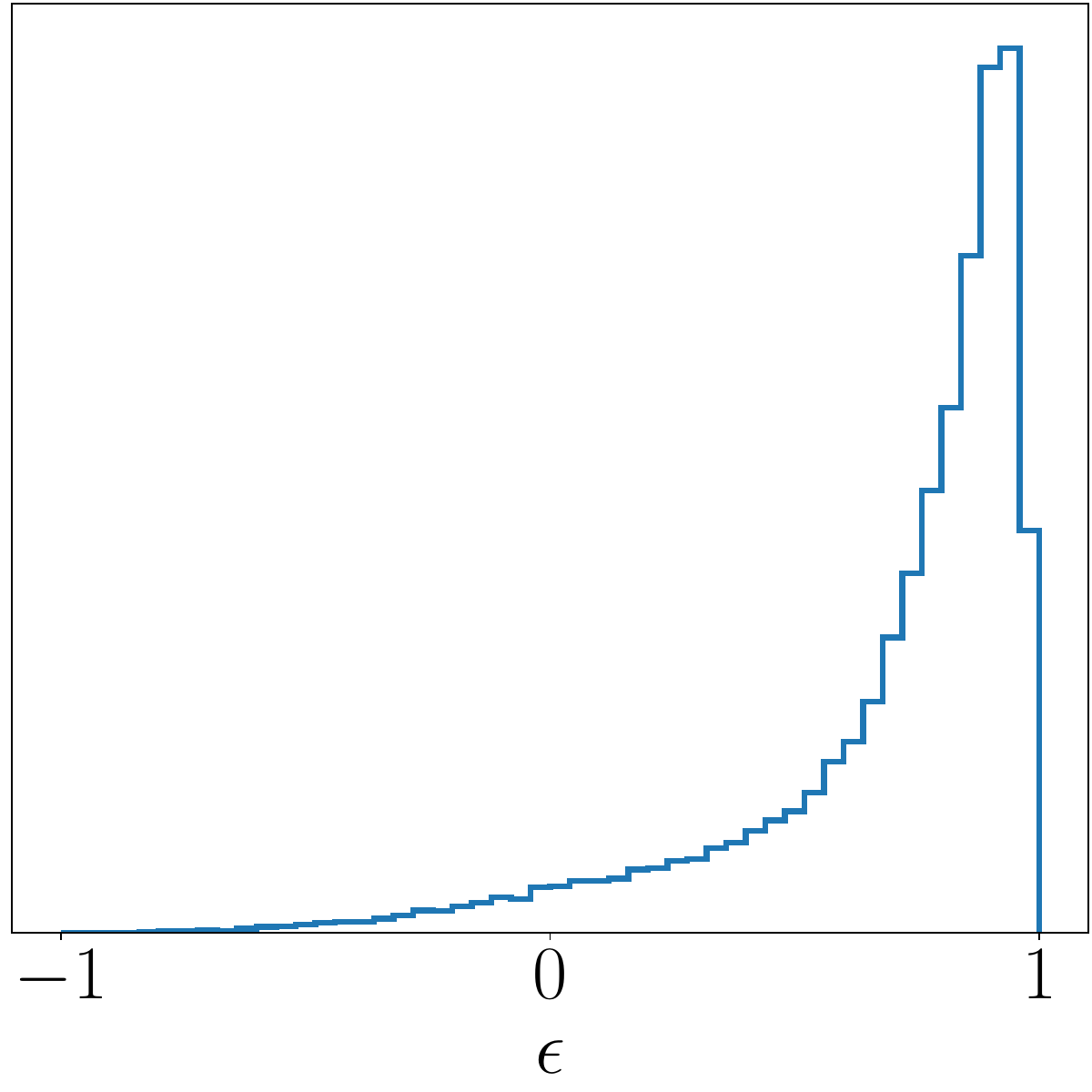}

%

  \caption{\label{fig_examples} Stellar mass density maps for galaxies with different morphologies in our samples (quoted in top panels). Top and middle rows correspond to face-on and edge-on views, with a green circle showing the galaxy radius, defined here as $3$ times the half mass radius of the stars. The size of each box is 50 kpc across. The bottom panels show, for each galaxy, the distribution of orbital circularities ($\epsilon$) of the stars. Galaxies in our mostly-disc sample have well-defined discs with a narrow distribution of $\epsilon$ close to 1 and a negligible contribution with $\epsilon \sim 0$ typical of spheroid-like components.}
 \end{center}
\end{figure*}

We use dynamical decomposition to quantify the morphology of the galaxies in the simulated sample. The dynamical decomposition method is based on the technique introduced in \citet{Abadi2003}, expanded to better take care of the assignment of particles to bulge/disc components in the central regions (Cristiani et al., in preparation). The method uses the distribution of circularities of the stars, $\epsilon$, defined as the angular momentum perpendicular to the discs divided by the angular momentum of a circular orbit with the same energy. A value $\epsilon =1$ indicates a star particle orbiting with a circular orbit perfectly aligned with the disc, and a value $\epsilon = -1$ indicates a star particle orbiting in a circular counter rotating orbit. $\epsilon =0$ corresponds to a purely radial orbit. 

Using the circularity distribution, the bulge or spheroid can be identified by mirroring the distribution of negative circularities, assumed by default to belong to the bulge, onto the positive $\epsilon$ axis to disentangle bulge from disc particles\footnote{An implementation of this and other dynamical decomposition methods is available at \url{https://github.com/vcristiani/galaxy-chop}} \citep[for details see][]{Abadi2003}. The mass associated to such $\epsilon$ distribution is then considered to be the mass in the spheroid for each galaxy, $M_{\rm sph}$. Note that, by default, this technique identifies non-rotating bulges, as it forces the distribution of circularities to be symmetric around $\epsilon = 0$. As such, the bulge definition is more aligned to the classical view of bulges as non-rotating and might not be well suited for rotating central components such as pseudo-bulges. Given the spatial resolution of the simulations, with a gravitational softening 740 pc, and the typical sizes of bulges $\sim 1$-$2$ kpc, the inner structure of this component remains mostly unresolved in the simulated galaxies in any case, making a simplified criteria such as the symmetric distribution of $\epsilon$ a good choice to differentiate a spheroidal component from the more rotationally supported discs.

Fig.~\ref{fig_RatioMass} shows the spheroid-to-total stellar mass ratio ($M_{sph}/M_{\star}$) as a function of stellar mass for all central galaxies considered in the simulation (gray dots). We select those with the least spheroidal component, $M_{sph}/M_{\star} < 0.12$, resulting on 43 mostly-disc galaxies (see blue symbols). This specific threshold is an arbitrary choice, since the $M_{sph}/M_{\star}$ distribution is continuous. We set the value such that the selected sample is sizeable in number and after visual confirmation that the criteria identifies well defined massive disc galaxies. We carefully select two additional comparison samples that follow the same distribution of $M_\star$ (according to a Kolmogorov-Smirnov test) but show different morphologies. First, $43$ galaxies in the intermediate sample (green symbols), defined as objects with $M_{sph}/M_{\star} \sim 0.3$. These intermediate galaxies represent still a population of disc-dominated objects, but with a more substantial contribution of the spheroid to the total mass. And second, we chose a population of $43$ spheroid-dominated galaxies, selected as those with the maximum $M_{sph}/M_{\star}$ at a given $M_\star$. Our total sample contains 129 (43 in each subsample) galaxies in the mass range: $\rm log(\rm M_\star/\mbox{M}_{\sun}) = [10.17-11.11]$. Table~\ref{tab_samples} lists the range of $M_{sph}/M_{\star}$ in each subsample.

Fig.~\ref{fig_examples} shows 5 examples of the galaxies included in our analysis. Top and middle rows correspond to the face-on and edge-on stellar projections \citep[made using Py-SPHViewer, ][]{benitez-llambay2015} and serve to illustrate the typical morphologies in our sample. Well defined and regular discs are common in our mostly-disc subsample (3 right-most columns) while the disc becomes weaker (second from left column) to not present (left column) for our intermediate and spheroid-dominated samples, respectively. This can be also clearly seen in the circularity distributions of these galaxies, shown in the bottom row. The spheroid-dominated galaxy has a circularity distribution peaking around $\epsilon = 0$ and no evidence of stars with large circularities, $\epsilon >0.7$, typically associated with a disc component. For the intermediate and mostly-disc examples, most of the mass (or area under the curve) is on orbits with high circularity, with the difference between these two types being the prominence of the ``bump" around $\epsilon =0$, associated to a bulge or spheroidal component. Encouragingly, some of our best examples in the mostly-disc subsample (see two rightmost panels in Fig.~\ref{fig_examples}) show narrowly peaked distributions around $\epsilon \geq 0.9$ and only a negligible fraction of particles with low $\epsilon$, indicating that almost pure-discs can and {\it do} form in $\Lambda$CDM simulations.

\begin{figure}
 \begin{center}
  \includegraphics[width=0.45\textwidth]{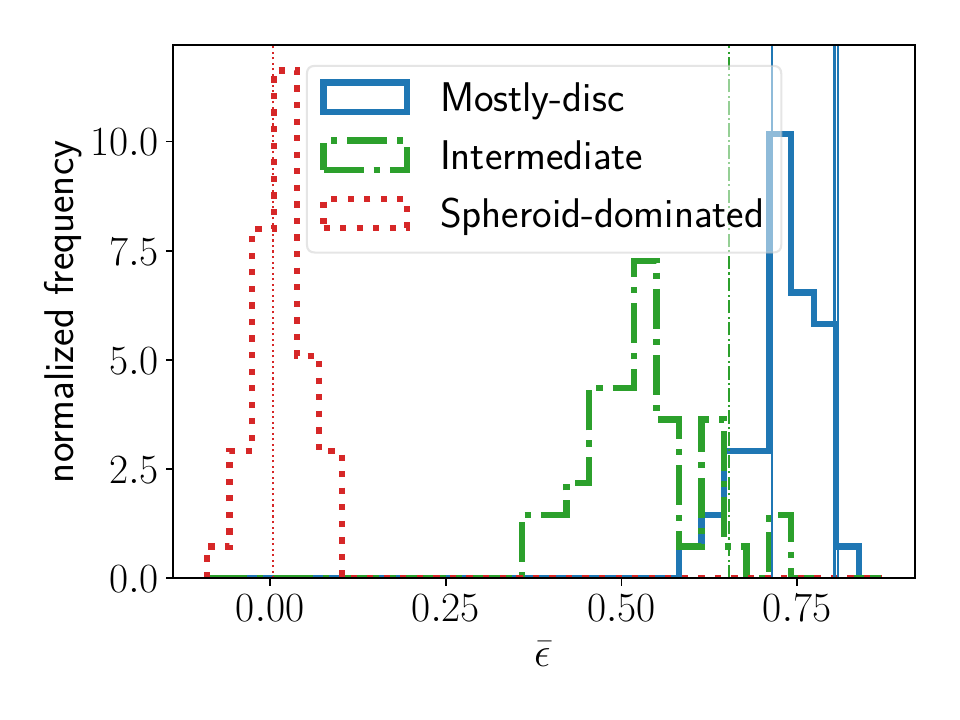}
  \caption{\label{fig_epsilon} Median value of the circularity parameter $\epsilon$ for the galaxies in our subsamples: mostly-disc (blue solid), intermediate (green dot-dashed) and spheroid-dominated (red dotted). We highlight with vertical lines the individual cases showcased in Fig.~\ref{fig_examples}. Our mostly-disc galaxies show median circularities $\epsilon \geq 0.7$ with $\sim 0.75$ the most typical value. 
  }
 \end{center}
\end{figure}

To characterise the morphology of our sample, we show in Fig.~\ref{fig_epsilon} the median circularity $\bar{\epsilon}$ of the stellar orbits in each simulated galaxy for our subsamples: mostly-discs (blue), intermediate (green) and spheroid-dominated (red). Thin vertical lines with the corresponding colours highlight where the examples shown in Fig.~\ref{fig_examples} lay in comparison to the rest of the subsample. All spheroid-dominated objects have $\bar{\epsilon} \sim 0$, consistent with being dispersion-dominated galaxies. The intermediate population has typically $\bar{\epsilon} \sim 0.5$, indicating the bi-modal distribution of mass into a low-$\epsilon$ component and a disc. Galaxies labelled as mostly-discs have $\bar{\epsilon} \sim 0.75$, which highlights the very large dominance of the disc as morphological feature. \citet{Peebles2020} correctly points out that $\epsilon \geq 0.7$, often used to define the disc component, may actually be too generous to describe observed pure-disc galaxies. However, for our specific sample, the baryonic treatment in TNG100 inherently pressurises the interstellar medium of gas at densities eligible for star-formation, being responsible for considerable ``thickening" of the vertical structure of the discs. From that perspective, the relatively high $\bar{\epsilon}$ found in our mostly-disc galaxy sample is reassuring and suggests that these objects may be used to shed light on the mechanisms able to build pure disc galaxies within the cosmological scenario.

\begin{table}
  \begin{center}
  \caption{
  \label{tab_samples} Range of the $M_{sph}/M_{\star}$ ratio for each sample.}
 \begin{tabular}{lc}
    \hline\hline
    Sample             & $M_{sph}/M_{\star}$ range\\\hline
    Mostly-disc        & 0.09-0.12\\
    Intermediate       & 0.29-0.32\\
    Spheroid-dominated & 0.87-1.00\\\hline
 \end{tabular}
 \end{center}
\end{table}

\begin{figure*}
 \begin{center}
  \includegraphics[width=0.45\textwidth]{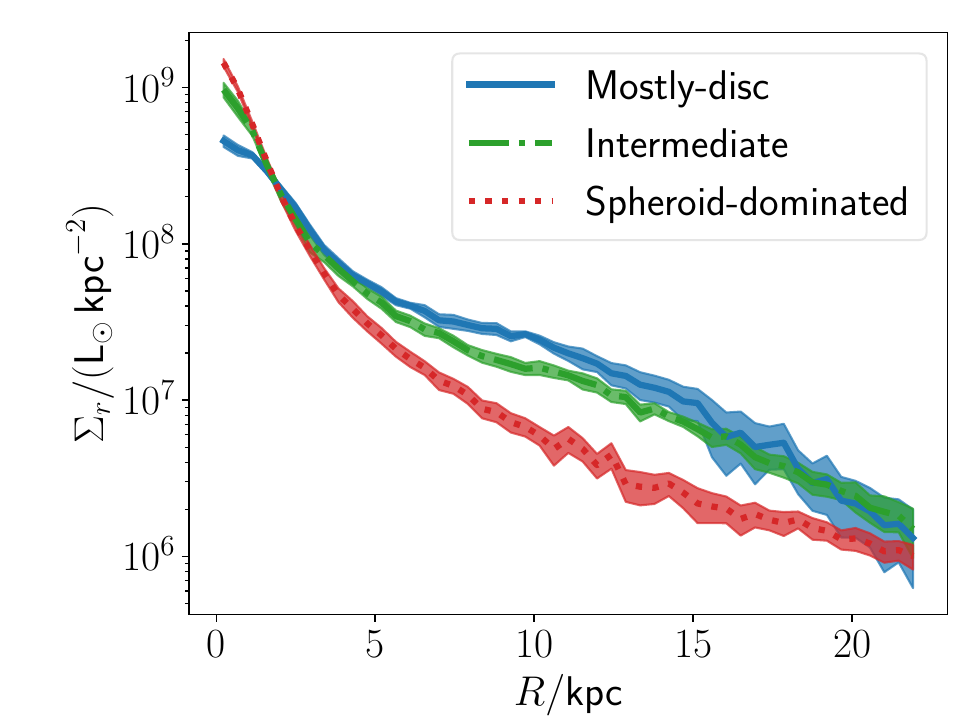}
  \includegraphics[width=0.45\textwidth]{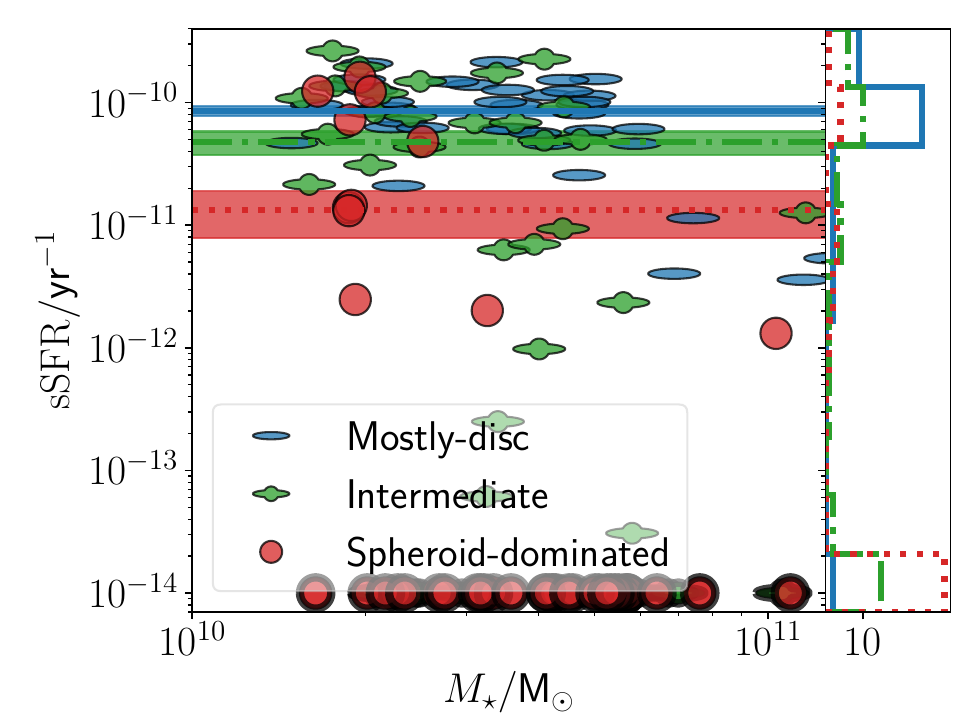}
  \caption{\label{fig_lumprofile} Luminosity profile as a function of projected radius (left) and specific star formation rate (sSFR) as a function of stellar mass (right) for galaxies in our samples. Thick lines indicate the median of each sample and shaded regions are the error on the median calculated from $500$ bootstrap resampling of the data. Our kinematics-based selection criteria reproduces expected traits from the different morphologies such as quiescence and steep luminosity profiles in spheroid-dominated objects vs. active star-formation and shallower luminosity profiles in disc-dominated samples.}
 \end{center}
\end{figure*}

\section{Assembly of stars, gas and dark matter in galaxies with different morphology}
\label{sec:assembly}

We examine in Fig.~\ref{fig_lumprofile} the $r$-band luminosity profile (left) and specific star formation rate (sSFR) as a function of $M_\star$ (right) for galaxies selected in our sample. Both diagnostics serve to differentiate galaxies with different morphologies: discs have exponential light profiles and form stars actively while spheroidal galaxies (or components) have cuspier light distributions consistent with de Vaucouleurs profiles and form little to no stars. In our sample, we find that the median light profile is indeed different for the subsamples, and consistent with these expectations mentioned above. Objects in the mostly-disc sample show a flatter light distribution and decreased inner concentration compared to the intermediate or spheroid-dominated galaxies. As mentioned above, lack of spatial resolution in the simulation prevents us from making a meaningful analysis of the profile slopes in the inner regions to determine whether they are more consistent with a classical-bulge or a pseudo-bulge. But we note that the central profile in our mostly-disc sample is substantially less concentrated and hints at a different slope than the spheroid-dominated median, suggesting that the processes shaping the inner light distribution in the mostly-disc sample and the spheroid-dominated sample are not the same.

The right panel in Fig.~\ref{fig_lumprofile} shows that, consistent with expectations, mostly-disc and intermediate galaxies are star forming while the majority of spheroids are quiescent \citep[assuming galaxies with $\rm sSFR < 10^{-11}\; \rm M_\odot/$ are quenched][]{Wetzel2013}. The median sSFR of the combined subsamples (horizontal lines) also follow expectations: mostly-disc galaxies have higher sSFR than intermediate and spheroids show the lowest values. While this average differentiation between star-forming and quiescent might have been expected based on our morphology selection, it is not a trivial exercise: galaxies were selected based only on a dynamical criteria for the orbital structure of stellar particles, yet, they are able to reproduce star-formation observables linked to gas properties, like the star formation main-sequence. This highlighting a clear success of this simulation and, more generally, of several current galaxy formation models able to reproduce a wide range of morphologies \citep[e.g. ][]{Vogelsberger2014, Schaye2015,Dubois2016}. We also note that there is an interesting overlap between a few individual objects of different subsamples, given by the large scatter in sSFR for a given galaxy type. For example, some low-mass spheroid-dominated galaxies are forming stars at rates comparable to the mostly-disc objects. Similarly, there are several mostly-disc and intermediate galaxies that are below the quenching threshold.

\begin{figure}
 \begin{center}
  \includegraphics[width=0.45\textwidth]{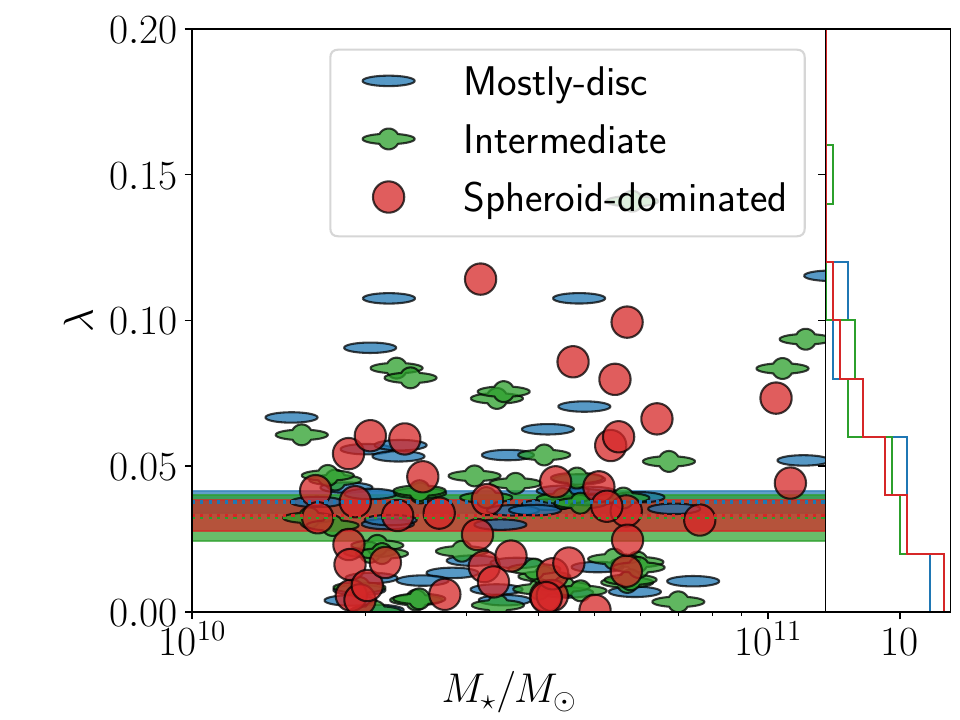}
  \caption{\label{fig_lambda} Halo spin, $\lambda$, as a function of stellar mass. Galaxies in all subsamples overlap in median (horizontal lines) and dispersion (symbols show individual objects in each sample), confirming that halo spin is not correlated with the morphology of the central galaxy in this mass range.
  }
 \end{center}
\end{figure}

With the aim to understand the link between the morphology of a galaxy and its dark matter halo, we compare next the halo spin parameter $\lambda$ of galaxies with different morphology. Fig.~\ref{fig_lambda}, shows that the distributions of halo spin for our mostly-disc (blue), intermediate (green) and spheroid-dominated (red) galaxies is indistinguishable from each other, reinforcing the idea that the halo spin is a poor predictor of morphology in the scale of MW-like galaxies \citep{Sales2012,Garrison-Kimmel2018,Rodriguez-Gomez2017}. Note that this is not the case for lower mass dwarfs, where discy morphologies are linked to the higher halo spins \citep{Rodriguez-Gomez2017,Benavides2023}.

\begin{figure}
 \begin{center}
  \includegraphics[width=0.45\textwidth]{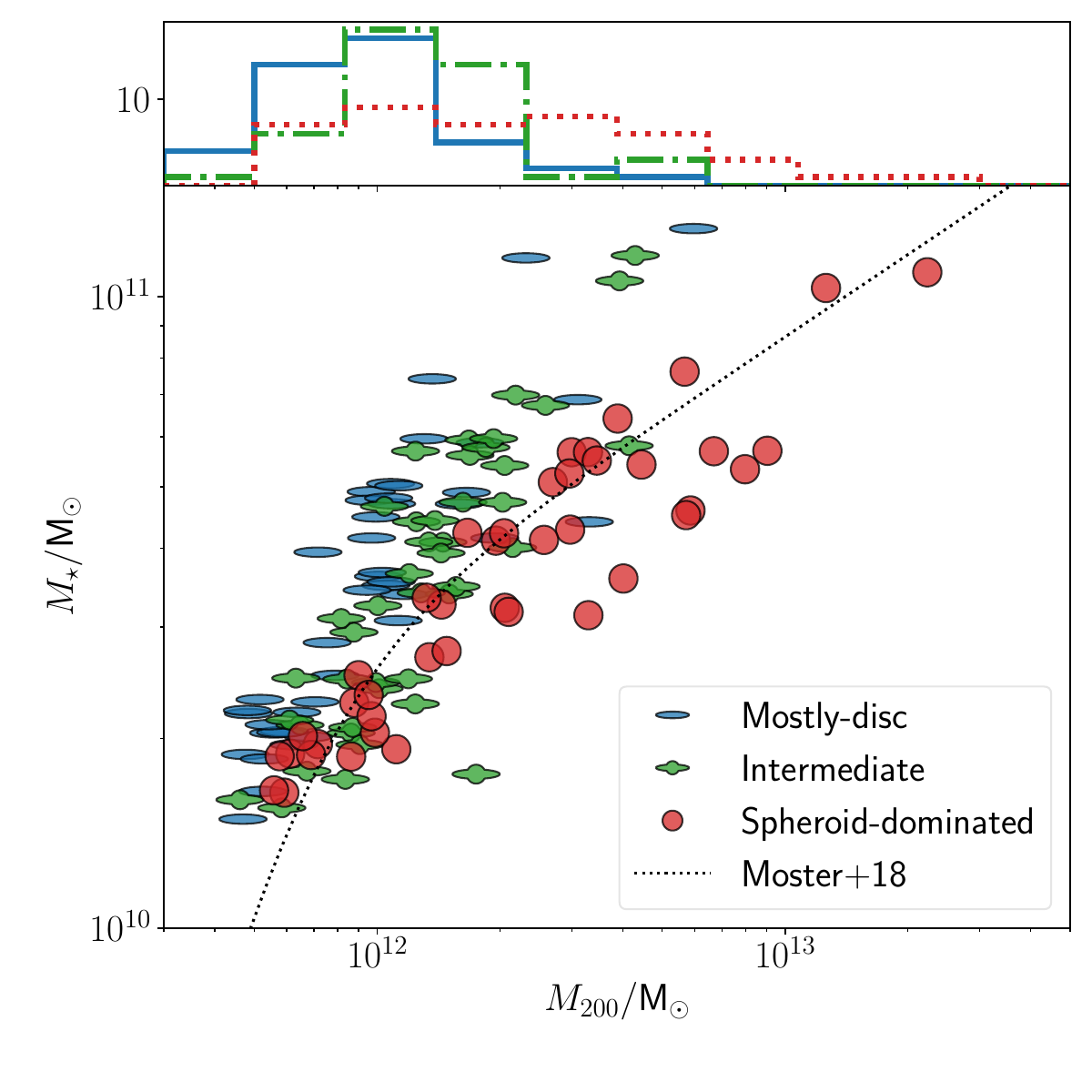}
  \caption{\label{fig_guo} Stellar mass - halo mass relation for the galaxies in our samples, indicated with symbols as before. To guide the eye, we also show the empirical fit by \citet[][]{Moster2018} in thin dotted line. There is a clear morphological bias in the relation, with disc-dominated objects populating haloes $2$-$10$ times lower mass than spheroid-dominated objects at the same $M_\star$. Similarly, at fixed $M_{200}$, disc-dominated galaxies form more stars than spheroids.}
 \end{center}
\end{figure}

We do, however, find a difference in the halo masses of galaxies with different morphologies. Fig.~\ref{fig_guo} shows the stellar mass - halo mass relation for objects in our sample (individual symbols, colour coded as before) along with the empirical fit from \citet{Moster2018} in dotted line for guidance. There is a clear bias in the distribution of galaxies, with mostly-discs having systematically lower-mass haloes than spheroid-dominated objects with the same $M_\star$. Intermediate galaxies fall in between. The bias is smaller for our lower $M_\star$ objects, but can reach halo masses $5$-$10$ times larger in the case of spheroid-dominated than mostly-discs for the brighter galaxies in our sample. Galaxies selected to have the same luminosity or stellar mass distribution (as in our case) may, therefore, greatly differ in the type of haloes they are sampling. This theoretical prediction is supported by several observational results suggesting more massive haloes around red/quiescent or spheroid-dominated galaxies at a given $M_*$ including the number of satellites \citep[e.g., ][]{Wang2012} and gravitational lensing studies \citep{Mandelbaum2016}.

Alternatively, results in Fig.~\ref{fig_guo} can be interpreted as discy galaxies being more efficient at transforming their available baryons into stars than spheroidals at a given halo mass. Partially, this view is supported by the previous discussion on the right panel of Fig.~\ref{fig_lumprofile}: mostly-disc galaxies continue to form stars actively until today, whereas the majority of spheroid-dominated objects are quiescent. The median stellar ages of the galaxies in our samples is $<\rm age> = 6.6 \pm 0.2$ Gyrs, $7.3 \pm 0.5$ Gyrs and $9.5 \pm 0.4$ Gyrs for the mostly-disc, intermediate and spheroid-dominated galaxies, respectively, showing that spheroids tend to stop their build up of new stars early on.

\begin{figure*}
 \begin{center}
    \includegraphics[width=0.95\textwidth]{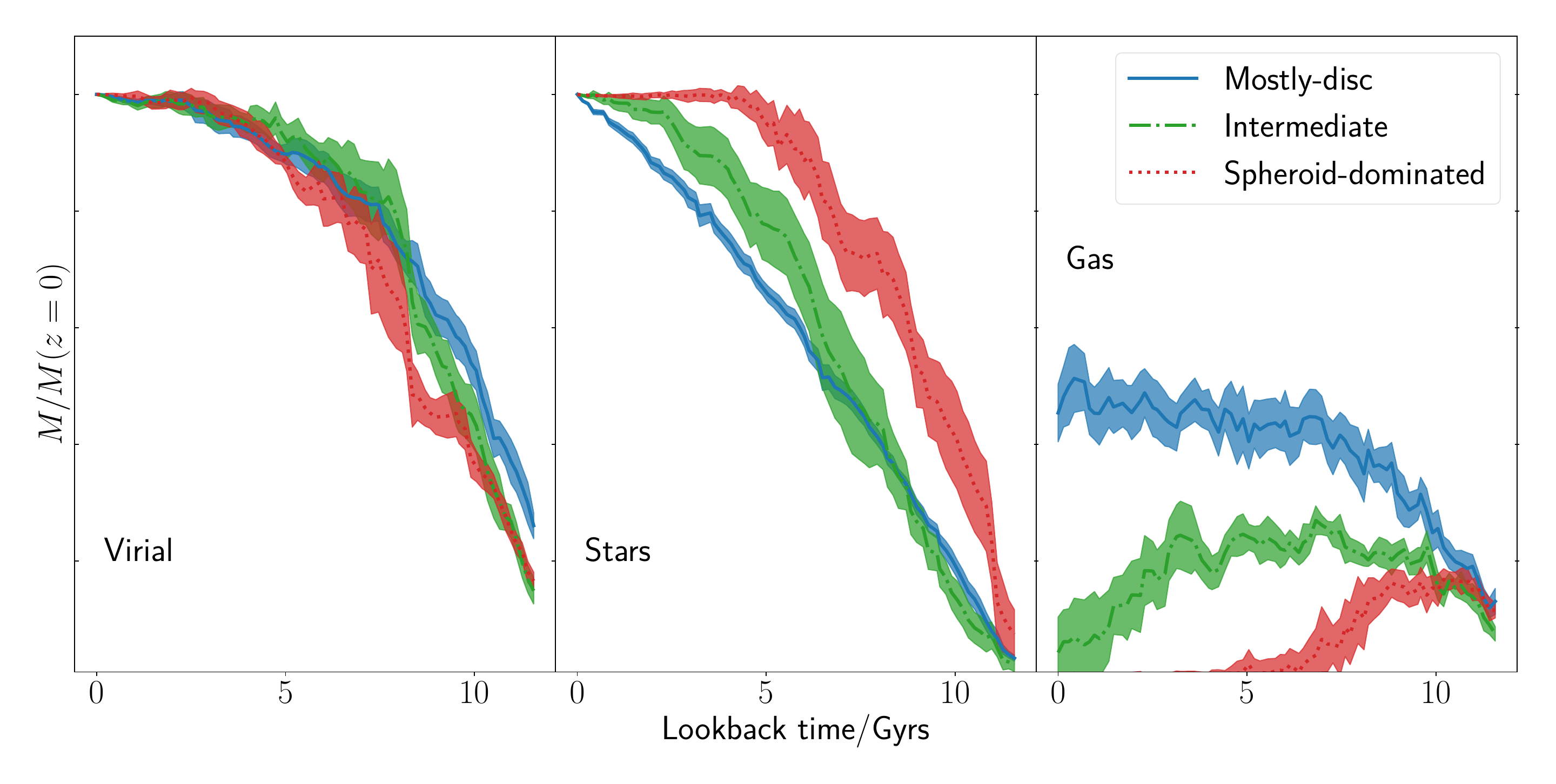}
    \caption{\label{fig_accretion} Assembly history for the virial mass (left), stellar mass (middle) and gas (right) components. Thick curves indicate the median of each subsample and the shaded region the error on the median calculated from $500$ bootstrap resampling of the data. Virial and stellar masses are normalised to their values at $z=0$ while the gas component is normalised to the stellar mass at $z=0$ for each subsample. Haloes grow at a similar pace between the samples, but stars and gas show clear differences: spheroid-dominated galaxies assemble their stars and consume their gas at earlier times than disc-dominated objects. Our mostly-disc sample shows steady star formation over time maintaining a richer gas supply than the intermediate and spheroid-dominated sample.
    }
 \end{center}
\end{figure*}

We explore this in more detail in Fig.~\ref{fig_accretion}, where we show the median assembly history of the halo mass (left), stars (middle) and gas (right) for each of our subsamples (shaded regions indicate the error in the median from 500 bootstrap sampling of the data). Curves are normalised to their final value at $z=0$ for each galaxy type in the case of halo mass and stars, and to the stellar mass at $z=0$ for the case of the gas (in which case shown values are indicative of gas fraction compared to stellar mass content today more than net gas mass at each time). The left panel indicates that the total virial mass grows, on median, similarly for all galaxies, regardless of their morphology, in particular for the late evolution.

Instead, the stellar mass behaves differently, with mostly-disc galaxies forming their stars substantially later than spheroids, and galaxies with intermediate morphology falling in between. Spheroid-dominated objects have formed or assemble stars in negligible amounts in the last $\sim 5$ Gyrs, which together with the similar evolution of the halo mass shown in the left panel helps explain their relative low stellar content compared to the halo mass shown in Fig.~\ref{fig_guo}. The gas evolution helps confirm this picture. Mostly-disc objects have sustained a healthy amount of gas  over time --about half that of their present-day stellar content-- in the last $\sim 8$ Gyr, while those destined-to-be spheroidals have only decreased their relative gas content since that time. Our intermediate morphology sample shows a behaviour somewhere in between, with some speed up in relative gas loss in the last $4$ Gyrs compared to the mostly-disc sample. It requires a healthy gas reservoir such as that in the mostly-disc galaxies to sustain an effective build up of stars that can place objects above the average stellar mass - halo mass relation. 

\section{The role of mergers}
\label{sec:mergers}

The relation between mergers and the resulting morphology is complex, often depending on the mass ratio of the merger, the orientation and gas content, as discussed in Sec.~\ref{sec:intro}. Here, we use the history of the main progenitors of our sample in addition to the stellar assembly catalogues \citep{Rodriguez-Gomez2015,stellarAssembly2, stellarAssembly3} provided in the Illustris-TNG database to quantify the contribution of mergers to the stellar mass build-up of galaxies with different morphologies. 

\begin{figure*}
\begin{center}
 \includegraphics[width=0.45\textwidth]{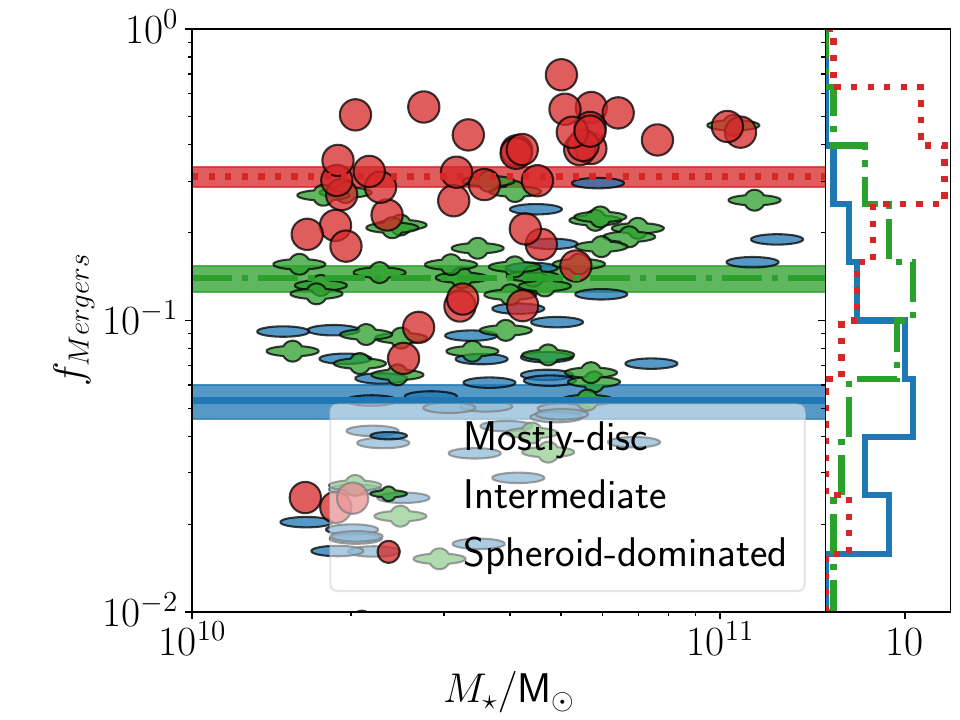}
 \includegraphics[width=0.45\textwidth]{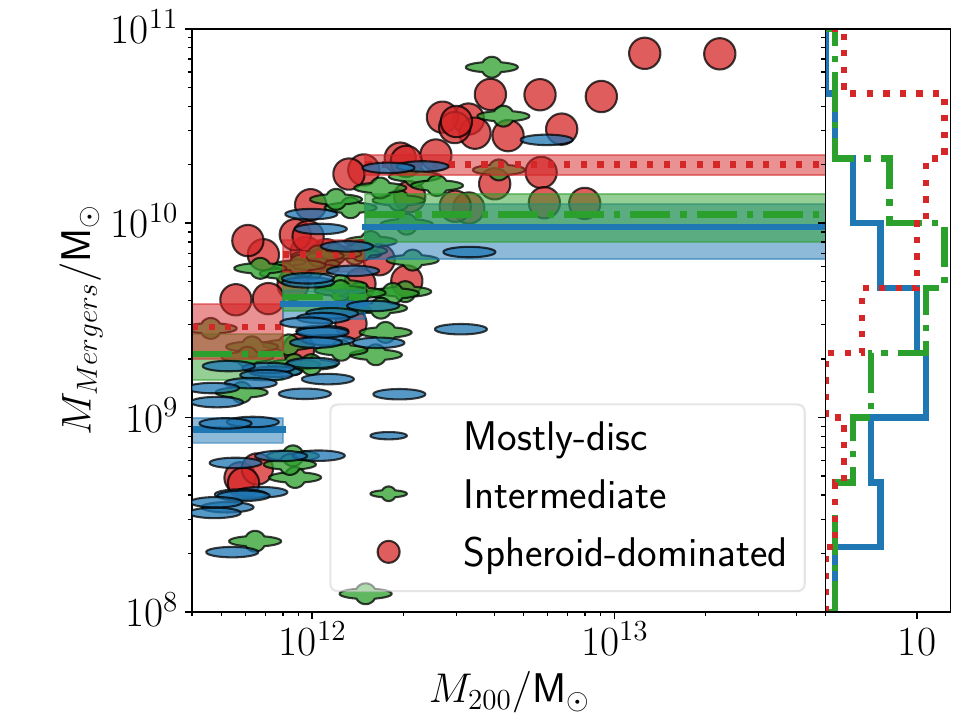}
 \caption{\label{fig_Fmerger} Fraction of the stellar mass from accreted origin as a function of stellar mass (left) and amount of stellar mass accreted as a function of virial halo mass (right) for our samples.  Symbols show individual objects, colour-coded as before, and horizontal lines show the median and error from bootstrap resampling. We show the medians calculated in 3 different $M_{200}$ ranges, ($<8 \times 10^{11}$), ($8 \times 10^{11} \rm - 1.5 \times 10^{12}$) and ($> 1.5 \times 10^{12}\; \rm M_\odot$) due to the biases in halo mass (see text for more detail). In general, spheroid-dominated objects have a larger contribution from accreted stars, but the overlap between the samples is substantial. We highlight the presence of mostly-disc and intermediate galaxies with a large accreted mass $M_{Mergers} > 10^{10}\; \rm M_\odot$ as well as the presence of 3 spheroid-dominated galaxies with less than $3\%$ of their mass from accretion. 
 }
\end{center}
\end{figure*}

Fig.~\ref{fig_Fmerger} shows that, at fixed stellar mass for the galaxy, the fraction of the stellar mass at $z=0$ that is of accreted origin is typically larger for spheroid-dominated than for our mostly-disc sample and intermediate galaxies. This follows traditional expectations of mergers building up dispersion-dominated components. The difference is substantial, with spheroids having, on median, $\sim 30\%$ of their stellar mass brought in by mergers, $\sim 12\%$ for the intermediate morphology and only $\sim 5\%$ of accreted stellar mass for our most rotationally supported discs. Shaded region around the median horizontal lines indicate the dispersion from $500$ bootstrap resampling of the data.

Two factors contribute to differentiate the accreted stellar mass fractions of spheroids and discs. First, the availability of gas to fuel in-situ star formation for longer times in the disc-dominated samples helps decrease the fraction of accreted stars by continuously building the in-situ component, process that is suppressed at later times in the spheroid-dominated sample. Second, the bias towards larger virial halo masses in the spheroid-dominated galaxies means that encounters with more massive satellites able to contribute larger amounts of stellar mass are more likely than in the disc-dominated samples. This is more clearly shown on the right panel of Fig.~\ref{fig_Fmerger}, where we show the total stellar mass of accreted origin as a function of $M_{200}$ instead of stellar mass. Once normalised to virial mass, the spheroid-dominated galaxies are on median only a factor $\sim 2$ more accreted mass than both disc-dominated samples and there is appreciable overlap of the samples.

In general, while the median trend indicates more prevalence of mergers in spheroid-dominated galaxies, there is a large scatter making the fraction of accreted mass or the total stellar mass accreted a poor predictor of morphology. For instance, there are three spheroid-dominated objects with $M_\star \sim 2 \times 10^{10}\; \rm M_\odot$ in our sample with only $\sim 2.5\%$ of their stellar mass of accreted origin, meaning that mergers had a negligible contribution to their dispersion-dominated structure \citep[see also ][]{Sales2012}. There is also a large degree of overlap between our mostly-disc and intermediate sample, suggesting that the processes leading to the formation of a sub-dominant bulge in mass are more complex than what it can be captured by these two simple merger statistics shown in Fig.~\ref{fig_Fmerger}.

Interestingly, several of our mostly-disc objects (blue) trace the upper end of the accreted mass at a given $M_{200}$, coexisting with the spheroid-dominated objects. In particular, $3$ out of the $5$ mostly-disc galaxies with $M_{200} \geq 1.5 \times 10^{12} \; \rm M_\odot$ have accreted $M_\star \geq 10^{10} \; \rm M_\odot$ and still retain their extreme discy kinematics. A closer inspection of some of the mostly-disc galaxies with large fraction of accreted masses reveals that the accreted stars show a wide distribution of circularities today, sometimes contributing to the ``bump" at $\epsilon \sim 0$ as expected, but in other cases showing intermediate circularities more reminiscent of a thick disc with $\epsilon \sim 0.5$ \citep{Abadi2003}. We conclude that for our mostly-disc galaxies showing substantial stellar accretion, what allows to retain their mostly-disc morphology is a strong in-situ component where the large majority of the stars is in nearly-circular orbits. This also emphasises the relevance of the additional efficiency in transforming gas into stars given their halo mass, as discussed in Fig.~\ref{fig_guo}.

\begin{figure*}
  \begin{center}
  \includegraphics[width=0.329\textwidth]{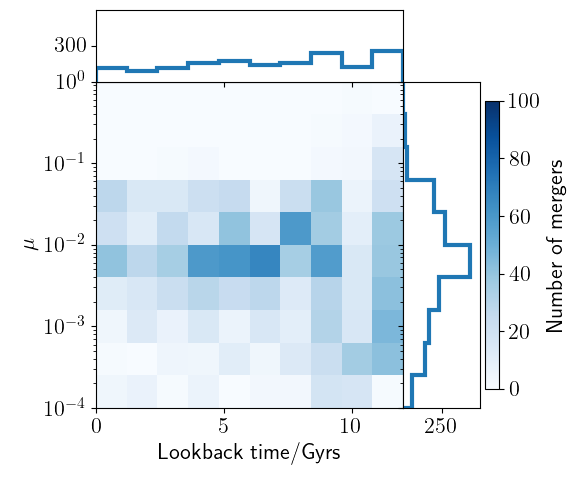}
  \includegraphics[width=0.329\textwidth]{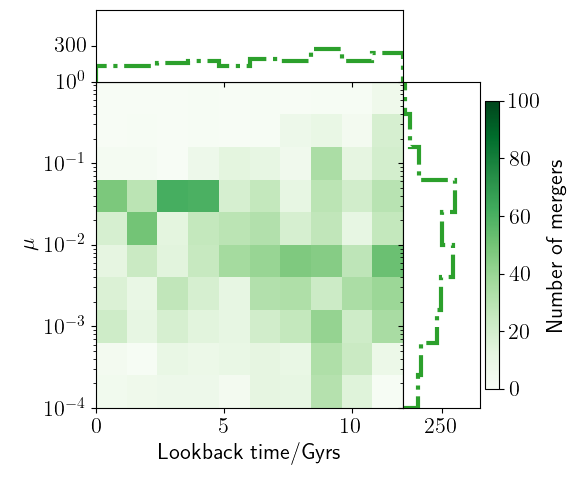}
  \includegraphics[width=0.329\textwidth]{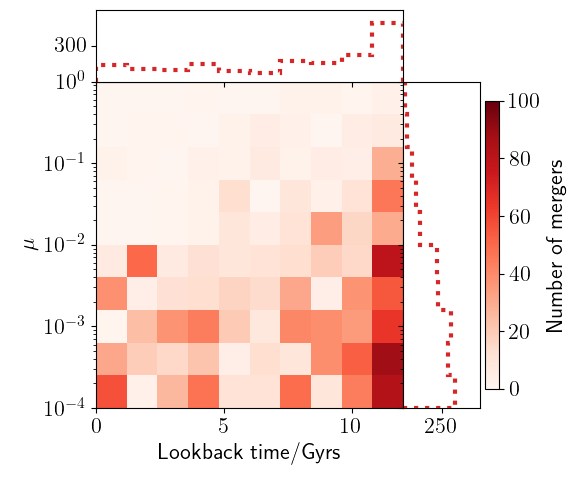}
  \caption{\label{fig_mergerTime} Distribution of the time (x-axis) and stellar mass ratio ($\mu$, y-axis) of the mergers for galaxies in each subsample: mostly-disc (left), intermediate (middle) and spheroid-dominated (right). Merger patterns are different across the galaxy types, with mostly-disc objects avoiding mergers in the last $\sim 4$ Gyrs compared to the other two subsamples. Merging seems also common for spheroid-dominated at very early times ($\geq 9$ Gyr ago). 
  }
 \end{center}
\end{figure*}

We also explore the timings of the mergers, finding interesting differences between the samples. Fig.~\ref{fig_mergerTime} quantifies the number of mergers at a given time and with a given stellar mass ratio, $\mu$, for each of our subsamples. To calculate this we use the main progenitors evolution after $z=4$ and measure the relative mass of each merger event, $\mu$, as the stellar mass ratio of the galaxies involved in the merger. For the main progenitor of our galaxies, we use their instantaneous stellar mass at the time of the merger. For the companion or merging galaxy, we follow the convention in \citep{Rodriguez-Gomez2015} and use their maximum mass prior to the merger to account for numerical and tidal stripping effects in $\mu$. 

Fig.~\ref{fig_mergerTime} shows that mostly-disc galaxies have a peak in merger activity $5$-$10$ Gyr ago, and the typical stellar mass ratio is 1:100 ($\mu \sim 10^ {-2}$). For intermediate morphology galaxies the mergers seem more recent, $t < 5$ Gyr ago, and the mass ratio is typically larger, with values in the range $\mu =[10^{-2} \rm - 10^{-1}]$ than in our mostly-disc sample. On the other hand, spheroid-dominated objects result from a more evenly-distributed set of mergers in time, although early ($t > 10$ Gyr) and recent ($t \sim 1$ Gyr) events seem common. The mass ratios involved in the spheroid-dominated sample are typically smaller than in the case of the disc-dominated subsamples, with $\mu \leq 10{-2}$ being the most typical. This highlights a clear prevalence of minor mergers over major-mergers in the build-up of our dispersion-dominated sample of spheroidal galaxies in our sample. 

\section{The settling and formation of the disc}
\label{sec:discform}

\begin{figure*}
\begin{center}
 \includegraphics[width=0.45\textwidth]{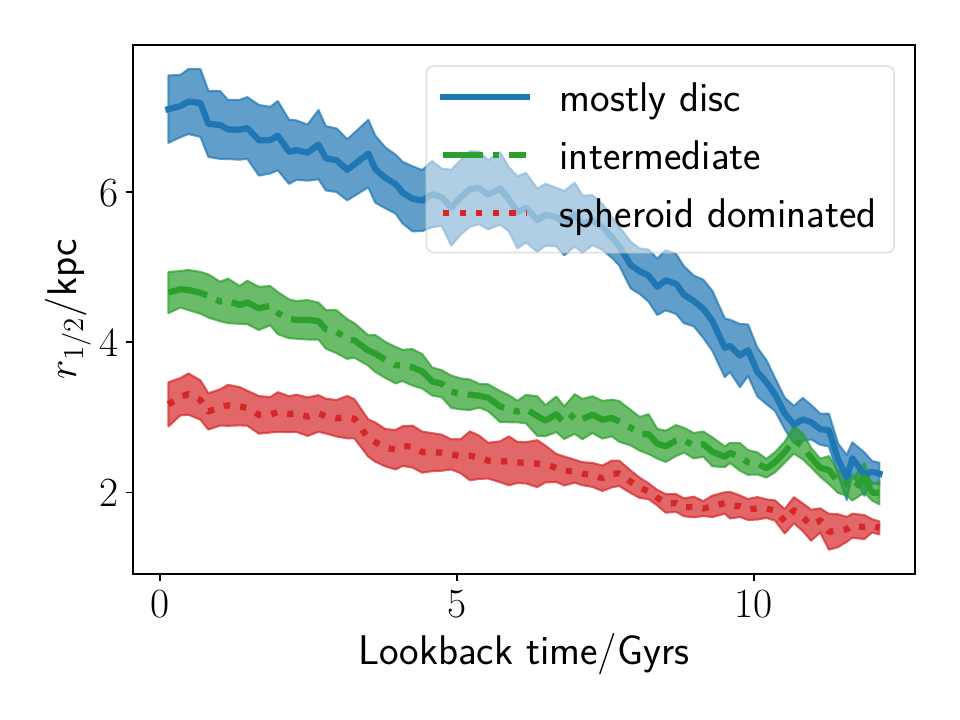}
  \includegraphics[width=0.45\textwidth]{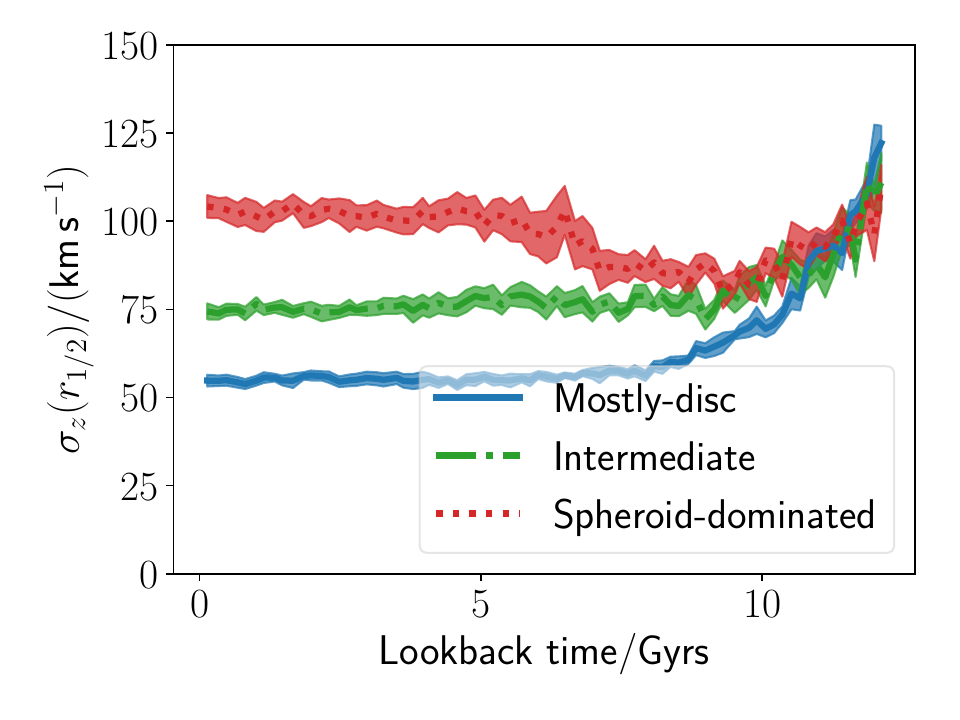}
 \caption{\label{fig_r5time} Structural evolution of the galaxies in our sample showing the median stellar half-mass radius, $r_{1/2}$, as a function of time on the left and the vertical stellar velocity dispersion measured at $r_{1/2}$ as a function of time on the right. Lines correspond to the medians of each subsample, shaded regions show the error following $500$ bootstrap resampling. Disc-dominated objects are more extended at all times compared to the spheroid sample. The velocity dispersion show similar value between the subsamples before $10$ Gyr ago, but they settle relatively quickly into their final values today around $8 \rm - 9$ Gyr ago, showing little evolution afterwards. 
 }
\end{center}
\end{figure*}

We now turn our attention to the time evolution of the structure in our samples. Fig.~\ref{fig_r5time} shows the stellar half-mass radius $r_{1/2}$ (left) and the stellar velocity dispersion in the z-direction at that radius $\sigma_z(r_{1/2})$ (right), as a function of time. Lines show the median of each subsample and the shaded region corresponds to the error calculated from $500$ bootstrap resampling. As expected, discs are today more extended and with less vertical velocity dispersion than the spheroids. The typical stellar half-mass radius in our mostly-disc galaxies is $\sim 6.5$ kpc, substantially more extended than galaxies like the Milky Way, and the typical vertical velocity dispersion at that radius is $\sigma_{z} \sim 50$ km/s. In comparison, the intermediate morphology and spheroidal samples show a more compact ($r_{1/2}\sim 4.5$ and $\sim 3.5$ kpc, respectively)  and dynamically hotter ($\sigma_z(r_{1/2}) \sim 75$ and $\sim 100$ km/s) stellar distribution today. 

In terms of sizes, Fig.~\ref{fig_r5time} suggests that the samples have mostly evolved parallel to each other, with the mostly-disc sample being larger than other galaxy types at all times, except perhaps for the very early evolution more than $10$ Gyr ago, when the mostly-disc and intermediate samples seem to become indistinguishable from each other. Spheroid-dominated objects are always more compact than those galaxies destined to be disc dominated. All galaxy types grow by factors $1.3 \rm - 2$ in size in the last $10$ Gyr. The mostly-disc galaxies show a steep size increase at intermediate times, $6$-$10$ Gyr ago, after which the size increase proceeds at a slower pace more similar to the other subsamples. The right panel of Fig.~\ref{fig_r5time} reveals an interesting behaviour: all galaxy types show similar $\sigma_z$ at early times, but start to differentiate $10$ Gyr ago, when the vertical velocity dispersion of the mostly-disc sample and the intermediate starts to decrease as a result of the disc formation. Remarkably, the average $\sigma_z$ of the mostly-disc objects reaches its present-day value of $\sigma_z = 52$ km/s very early on, staying relatively constant in the last $\sim 9$ Gyr.

\begin{figure*}
 \begin{center}
  \includegraphics[width=0.45\textwidth]{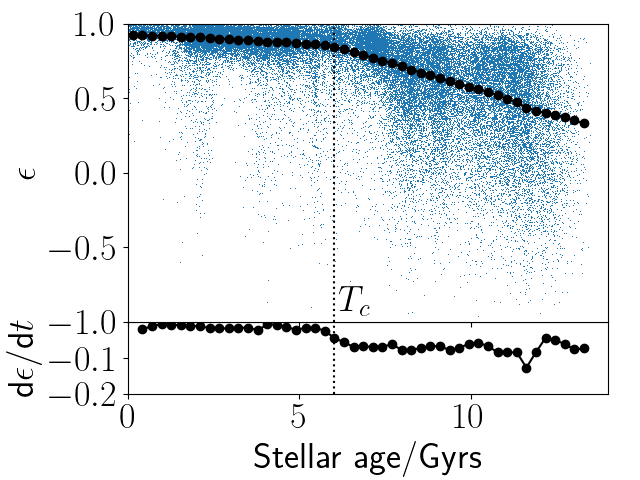}
  \includegraphics[width=0.45\textwidth]{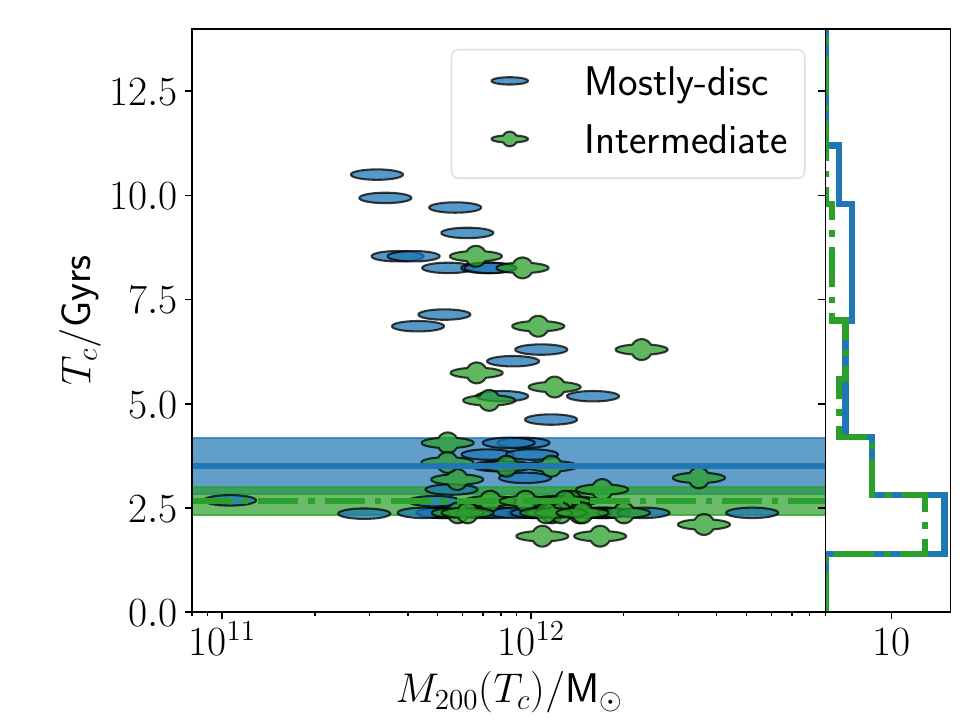}
  \caption{\label{fig_tcdef} Formation time of the discs in our sample. The left panel shows an example of our definition based on the circularity parameter as a function of the stellar age of all stellar particles in a given galaxy (blue dots). We calculate the running median (black symbols) and their derivative (bottom panel) to find the time when dominant and consistent formation of stars in circular orbits occurs, $T_c$. On the right panel we show $T_c$ for all disc-dominated galaxies as a function of their virial mass at the time of $T_c$ (individual symbols). On median, mostly-disc objects settle into circular orbits slightly earlier than for the intermediate objects, but the scatter is large. Galaxies reach $T_c$ typically late and when their host halo mass is larger than $M_{200} \sim 3 \times 10^{11}\; \rm M_\odot$ (with one exception). Some of the oldest discs have been consistently forming stars since $\sim 10$ Gyr ago.
  }
 \end{center}
\end{figure*}

To better identify the formation time of individual discs, we look into the orbital structure of the stars in our simulated galaxies today and determine the stellar age at which stars started to form consistently aligned and in circular orbits. The left panel of Fig.~\ref{fig_tcdef} shows an illustration of the procedure. We start with the circularity $\epsilon$ as a function of stellar age in the main panel for all the stars in one of our mostly-disc galaxies. For the older stars there is a wide range of $\epsilon$ which becomes increasingly more concentrated towards $\epsilon \sim 1$, corresponding to circular orbits, for younger stars. The black curve shows the median as a function of stellar age, applying a moving average smoothing with $dt = 4.2$ Gyr \citep[Fig. 5 in][shows a similar analysis but using \lbrack Fe/H\rbrack\ instead of the stellar age]{Chandra2023}. Next, we calculate the time derivative of this median (bottom left panel) and we calculate the mean and dispersion at those times using all the values of $\mbox{d}\epsilon/\mbox{d}t$ that correspond to a given age or younger. Lastly, we measure the first time that the value of $\mbox{d}\epsilon/\mbox{d}t$ is smaller that the aforementioned mean minus $n$ times the dispersion. We call this time $T_c$, for time of consistent alignment in the stars. After visual inspection of several examples, we find that a value of $n=2$ is suitable for our objectives. In other words, $T_c$ is the smaller value that comply with:

\begin{equation}
 \mbox{d}\epsilon(T_c)/\mbox{d}t < M(\mbox{d}\epsilon/\mbox{d}t, T_c) - n \times S(\mbox{d}\epsilon/\mbox{d}t, T_c)
\end{equation}

\noindent
where $M(\mbox{d}\epsilon/\mbox{d}t, T_c)$ is the average value up to $T_c$, and $S(\mbox{d}\epsilon/\mbox{d}t, T_c)$ is the dispersion up to $T_c$. To gain some feeling for what $T_c$ measures, we can also consider the median circularity of the stellar orbits measured at $T_c$. For the specific case shown in the left panel of Fig.~\ref{fig_tcdef}, we find $\epsilon(T_c) = 0.87$. More generally, we find the median of $\epsilon(T_c)$ is $0.81\pm0.02$  and $0.83\pm0.02$, for our mostly-disc and intermediate sample, highlighting the power of $T_c$ to identify the formation time of discs dominated by circular orbits. Note that alternative measurements to the formation time of discs have been introduced in the literature before, including, for example, the ``spin-up" time, designed to capture the moment at which the thin-disc starts to form \citep{Belokurov2022,Yu2023, Chandra2023}. While our $T_c$ is related to certain degree to the spin-up time of discs, we favour $T_c$ instead for our study as a more general property to characterise the build-up of rotationally supported discs in a wide range of masses and not only focused on MW-like galaxies.

The right panel in Fig.~\ref{fig_tcdef} shows the distribution of $T_c$ for both of our disc-dominated subsamples as a function of the virial mass of the haloes at that time\footnote{there are 11 galaxies for which the $T_c$ condition is never met, 1 from the mostly-disc sample and 10 with intermediate morphology, which have all been removed in Fig.~\ref{fig_tcdef}}. On median, the stars in mostly-disc galaxies become consistently aligned $3.9 \pm 0.7$ Gyrs ago, with the intermediate sample showing slightly more recent values ($2.9 \pm 0.3$ Gyr). However, there is a large dispersion in the samples, and several of the mostly-disc galaxies have $T_c > 7$ Gyr. Considering the many factors threatening disc survival in $\Lambda$CDM, it is remarkable that these objects manage to continuously build a disc for $7$-$10$ Gyr. In general, stellar orbits seem to begin settling when the halo mass reaches $M_{200} \geq 3 \times 10^{11}\; \rm M_\odot$, in agreement with several recent studies on disc formation \citep{Yu2023,Chandra2023}. Such scale might be associated with the inception of a hot circum-galactic medium \citep[e.g., ][]{Jahn2022} which facilitates the accretion of gas into the disc coherently aligned. The important role of hot accretion for the build-up of discs was highlighted in \citet{Sales2012} and has been confirmed in other simulations and models \citep[e.g., ][]{Ubler2014,Hafen2022,Yu2023,Stern2024,Semenov2024}.

\begin{figure}
 \begin{center}
  \includegraphics[width=0.45\textwidth]{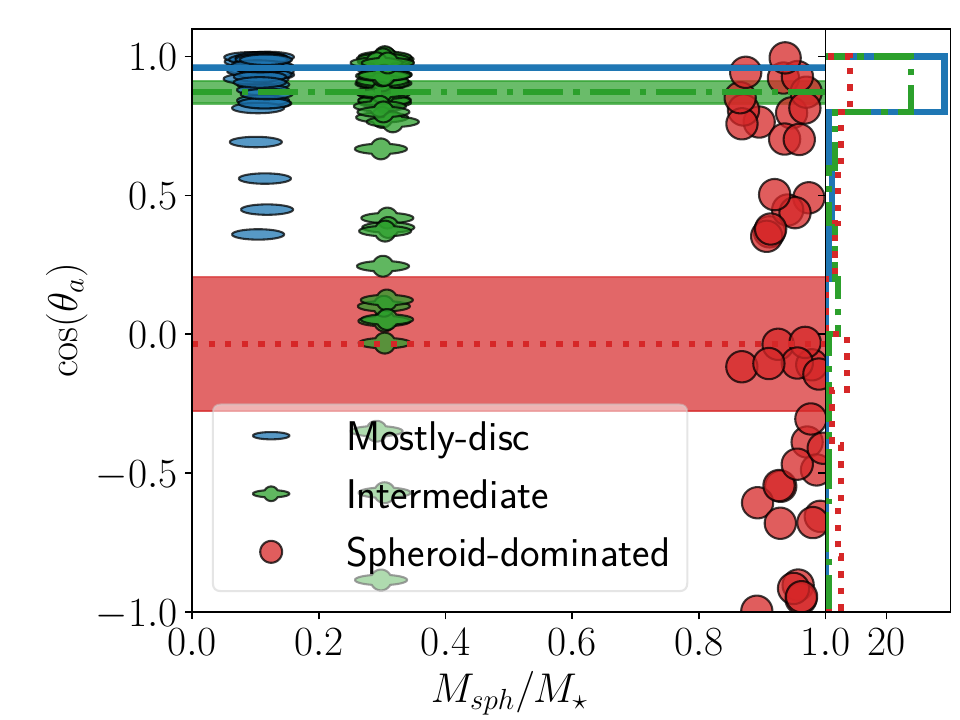}
  \caption{\label{fig_alig} Angle between the angular momentum of the gas in the halo and the stars in the galaxy, $\theta_{a}$, as a function of the mass in the spheroidal component, $M_{sph}/M_{\star}$, for galaxies in our sample. We show this at $T_c$ for both disc-dominated samples and use $z=0$ for spheroid-dominated galaxies. Medians are highlighted with horizontal lines as before. Mostly-disc galaxies show a remarkable coherently aligned gas angular momentum with respect to the one of the stars already formed at $T_c$, with a median $\rm cos(\theta_{a}) \sim 0.95$) and a few less well-aligned objects. The intermediate sample shows also a preference for strong co-rotation of the components ($\rm cos(\theta_{a}) > 0.5$), but the angle between the angular momentum of gas and stars can be larger, including counter-rotation for a few objects. Spheroid-dominated are here for comparison only and show a wide range of misalignment.
  }
 \end{center}
\end{figure}

We measure the alignment of gas at $T_c$ for the different subsamples in Fig.~\ref{fig_alig}. For each galaxy, we select the data of the main progenitor from the snapshot closer to $T_c$ (or at $z=0$ for the spheroid-dominated sample and for the galaxies where we were unable to calculate $T_c$), and calculate the angular momentum of the gas that is part of the host halo up to $r_{200}$ (without considering gas associated to satellites, if any). We measure  $\theta_{a}$ as the angle between the angular momentum of this gas and the angular momentum of the stars in the galaxy. Fig.~\ref{fig_alig} shows the values of $\cos(\theta_{a})$ as a function of the ratio between the mass of the spheroidal component, $M_{sph}$ and the total galaxy mass $M_{\star}$ (this is for convenience to avoid overlap in the samples). Disc-dominated galaxies in general show a better alignment than the spheroid-dominated objects, highlighting the importance of gas alignment in the build up of discs. Even within the disc-dominated galaxies, those with mostly-disc morphology at $z=0$ show even a tighter alignment than the intermediate galaxies, agreeing with expectations that misaligned gas accretion may correlate with the formation and growth of spheroidal components \citep{Scannapieco2009,Sales2012}.

\section{Summary and Conclusion}
\label{sec:concl}

We take advantage of the large volume simulated within the Illustris TNG100 simulations to study the formation and evolution of the most rotationally supported galaxies within $\Lambda$CDM. Using dynamical decomposition to identify spheroid- and disc- components, we find a population of galaxies with negligible ($\leq 12\%$) amount of stars on spheroid-like orbits. Numerical limitations prevent us from assessing whether the small bulge component is consistent with a classical vs. pseudo-bulge morphology. Nevertheless, the formation of such mostly-disc galaxies can cast some light on the processes necessary to form purely-disc spirals in the Universe and how they can exist within the cosmological scenario. 

Mostly-disc morphologies arise as a combination of very efficient star formation given a halo mass, probably as a consequence of a stable and abundant supply of gas maintained during the last $\sim 10$ Gyr, in addition to an earlier settling of the stellar orbits into circular or quasi-circular motions supported by a well-aligned distribution of angular momentum between the existing stars and the infalling gas. They also typically avoid mergers in the last $\sim 5$ Gyr. At a given halo mass, disc-dominated galaxies are $\sim \times 2$ more massive in stars than spheroidal-dominated objects with the same halo mass, placing discy morphologies preferentially above the expected abundance matching relation. On average mergers play a more important role in the build-up of spheroid-dominated galaxies, but it is possible to find mostly-disc and intermediate morphology galaxies with equal or larger amount of accreted stellar mass than spheroidals. Similarly, a few of our spheroid-dominated galaxies are formed with a negligible contribution from mergers. 

Overall, we do not identify any ``special" mechanism that distinguishes the formation of our mostly-disc galaxies from the (still disc-dominated) intermediate morphologies, but instead, we find that the mostly-disc sample is always part of a continuous distribution with slightly more efficient star formation, more and better aligned gas supply and earlier settling of the disc than objects with more massive bulge components. If an extrapolation of these results can be used to interpret the formation of bulgeless galaxies, one interesting question to be addressed is why such conditions would be met so easily in the Local Volume, where more than half of the large spirals show a negligible bulge component \citep{Kormendy2010,Peebles2020}. Alternatively, a more dramatic interpretation is that our current understanding of $\Lambda$CDM in combination with galaxy formation models are still unable to properly predict the morphology of galaxies and, in particular, we are lacking physical processes that enhance the rotational support of the stars, in particular in the central regions of galaxies.

\begin{acknowledgements}
SR, VC and MA acknowledge financial support from FONCYT through PICT 2019-1600 grant. LVS acknowledge financial support from NSF-CAREER-1945310 and NSF-AST-2107993 grants.
\end{acknowledgements}

\bibliographystyle{aa}
\bibliography{biblio}


\end{document}